\documentclass[11pt,a4paper,notitlepage]{article}
\usepackage[utf8]{inputenc}
\usepackage[english]{babel}
\usepackage[T1]{fontenc}
\usepackage{hyperref}
\usepackage{amsmath}
\usepackage{stmaryrd}
\usepackage{amsfonts}
\usepackage{amssymb}
\usepackage{color} 
\usepackage{caption}
\usepackage{amsmath}
\usepackage{relsize,exscale}
\usepackage{array,multirow,makecell}
\usepackage[toc,page]{appendix}
\DeclareSymbolFont{largesymbols}{OMX}{cmex}{m}{n}

\setcellgapes{1pt}
\makegapedcells
\newcolumntype{R}[1]{>{\raggedleft\arraybackslash }b{#1}}
\newcolumntype{L}[1]{>{\raggedright\arraybackslash }b{#1}}
\newcolumntype{C}[1]{>{\centering\arraybackslash }b{#1}}

\newtheorem{theorem}{Theorem}
\newtheorem{definition}{Definition}
\newtheorem{property}{Property}
\newtheorem{proposition}{Proposition}

\newtheorem{corollary}{Corollary}

\newcommand{\sym}{\mathrm{Sym}}

\newtheorem{lemma}{Lemma}
\usepackage{enumerate}
\usepackage{subfig}
\usepackage{graphicx}

\newcommand{\cG}{{\mathcal G}}

\newcommand{\droit}{\mathrm{d}}
\newcommand{\SU}{\mathrm{SU}}

\newcommand{\beq}{\begin{equation}}
\newcommand{\eeq}{\end{equation}}
\newcommand{\bea}{\begin{eqnarray}}
\newcommand{\eea}{\end{eqnarray}}
\definecolor{mygray}{gray}{0.3}

\newcommand{\bes}{\begin{eqnarray}}
\newcommand{\ees}{\end{eqnarray}}

\newcommand\restr[2]{{
  \left.\kern-\nulldelimiterspace 
  #1 
  \vphantom{\big|} 
  \right|_{#2} 
  }}

\newcommand{\U}{\mathrm{U}}
\usepackage{multicol}
\usepackage{amssymb}

\makeatletter
\begin{document}
\begin{center}
\textbf{\Large{Ward-constrained melonic renormalization group flow for the rank-four $\phi^6$ tensorial group field theory}}\\
\medskip
\vspace{15pt}

{\large Vincent Lahoche$^a$\footnote{vincent.lahoche@cea.fr}  \,\,and 
Dine Ousmane Samary$^{a,b}$\footnote{dine.ousmanesamary@cipma.uac.bj}
 }
\vspace{15pt}

a)\,  Commissariat à l'\'Energie Atomique (CEA, LIST),
 8 Avenue de la Vauve, 91120 Palaiseau, France

b)\, Facult\'e des Sciences et Techniques (ICMPA-UNESCO Chair),
Universit\'e d'Abomey-
Calavi, 072 BP 50, B\'enin
\vspace{0.5cm}
\end{center}
\begin{center}
\textbf{Abstract}
\end{center}
The nontrivial fixed point discovered for $\phi^4$-marginal couplings in tensorial group field theories have been showed to be incompatible with Ward-Takahashi identities. In a previous analysis, we have stated that the case of models with interactions of order greater than four could probably lead to a fixed point compatible with local Ward's identities. In this paper, we focus on a rank-$4$ Abelian $\phi^6$-just renormalizable tensorial group field theory and describe the renormalization group flow over the sub-theory space where Ward constraint is satisfied along with the flow, by using an improved version of the effective vertex expansion. We show that this model exhibit nontrivial fixed points in this constrained subspace. Finally, the well-known asymptotically freedom of this model is highlighted.


\section{Introduction}

Group field theories (GFTs) are a type of non-local fields theories defined on $d$-copies of a group manifold. From more than one decade, they have been considered as a promising way to quantize gravity \cite{Pagani:2019vfm}-\cite{Oriti:2009wn}. A well-defined theory for quantum gravity is necessary to understand the nature of space, time and geometry, and especially to address the questions about the origins of our universe. Indeed, the current description remains incomplete and fail at the Planck scale $\ell_p=\sqrt{G\hbar/c^3}$, because of the incompatibility between general relativity (GR) and quantum theory (QT) \cite{Ashtekar:2014kba}. In the last years, some important developments are given in various directions to think about the question of quantum gravity, such as random geometry, canonical quantum gravity, and covariant approach with spin-foam models, which have together converged toward the definition of GFTs \cite{Rovelli:1997yv}-\cite{Aastrup:2006ib}. In particular, GFTs provide a unified and convenient field theoretical framework to discuss the second quantization of loop quantum gravity (LQG) states, and their Feynman amplitudes coincide with spin-foam amplitudes, with a canonical weight \cite{Perez:2002vg}-\cite{Colosi:2004vw}. Tensorial group field theories (TGFTs) is an improvement of the standard GFTs, including tensoriality as a strong criterion to build interactions \cite{Gurau:2011xq}-\cite{DiFrancesco:1993cyw}. Tensoriality comes from colored random tensors models (RTMs) and corresponds to a specific invariance concerning internal unitary transformations \cite{Gurau:2011xq}-\cite{DiFrancesco:1993cyw}. RTMs provide a nice generalization of random matrix models; and are viewed as a convenient formalism for studying random geometry for dimensions higher than two \cite{DiFrancesco:1993cyw}. The first success of RTMs was the discovery by Gurau in (2009) of a $1/N$ expansion for tensors, analogous to the t' Hooft expansion for matrix models from which the genus being replaced by a non-topological invariant called Gurau's degree \cite{Gurau:2011xq}-\cite{Gurau:2010ba},\cite{DiFrancesco:1993cyw}. Tensoriality have pointed out to be a very strong improvement, allowing to define renormalization group and renormalizable actions. Renormalization and renormalizability take place into a \textit{geometrogenesis scenario} for space-time emergence \cite{BenGeloun:2011rc}-\cite{Carrozza:2013wda}. In this scenario, the large scale structure of space-time concerning the Planck scale is described as a GFT condensate, analogous to the Bose-Einstein condensates in bosonic many-body systems \cite{deCesare:2016rsf}-\cite{Gielen:2018xph}. Renormalization group (RG) allows to build effective field theories from an elementary scale, and to understand dynamical phase transitions in the statistical and quantum system (see \cite{Wilson:1971bg}-\cite{Lahoche:2018vun} and the references therein). With this respect, it was considered as a very important tool to understand the condensation mechanism in GFT. \\

\noindent
The RG equations describe how the effective actions move when quantum fluctuations higher than a reference running scale are integrated out, building a path from the deep UV scale toward the large IR scales
\cite{Wilson:1971bg}-\cite{Wetterich:2001kra}. The trajectories from UV to IR are described as a flow through the infinite-dimensional functional space of allowed actions, the theory space. Nonperturbative methods to build the RG group exist in standard field theories, the most popular and tractable being the functional renormalization group (FRG) method, based on the Wetterich-Morris equation \cite{Wetterich:1992yh}-\cite{Wetterich:2001kra}. These standard tools have been successfully applied for TGFT, a success especially due to the flexibility and the simplicity of the formalism, allowing to deals with the specific non-localities of the TGFT interactions. Nonperturbative RG equations are known to be difficult to solve for TGFT models, as well as in standard field theory. This is why a large part of the studies on this topic limit their investigation of the truncation approximation. Truncation consists of a systematic projection into a reduced dimensional phase space along which the flow equations may be solved analytically or numerically. This methods has been performed for a very large class of models \cite{Geloun:2016qyb}-\cite{BenGeloun:2018ekd}. Interestingly, all of them reveal the occurrence of a nonperturbative fixed point in the \textit{symmetric phase}, that is, as long as the vanishing mean-field remains a good vacuum around which we can expand the flow equation. These fixed points play different roles, but are crucial, both for UV consistency of the theory and to support the condensation scenario. For the first case, UV fixed point ensure the UV completion of the models, which are stated to be asymptotically free and (in the worst cases asymptotically safe \cite{Carrozza:2016tih}). IR fixed point, in contrast, corresponds to non-trivial sums of spin-foam amplitudes at large scale, and they are reminiscent of a phase transition behaviour. \\

\noindent
In the deep UV, instead of a crude truncation into the full theory space, a recent approach \cite{Lahoche:2019vzy}-\cite{Lahoche:2018vun} propose to solve the same question, by replacing the truncation by the effective vertex expansion (EVE) in the reduced phase space with a closure of the hierarchical system derived from the exact RG group equation. To be more precise, in the EVE approach, we use of relevant and marginal coupling as drivers of the full RG flow in the deep UV and solve the RG equations discarding all the irrelevant contributions. The resulting equations, moreover, keep the full momenta dependence of the vertex function, allowing to investigate beyond the local potential approximation. This, in particular, plays an important role in the derivation of the anomalous dimension and provides a relevant correction concerning the truncations. In addition to these closure relations, the Ward Takahashi identities provide some constraints which have to be solved simultaneously with the flow equations, at the same level of approximation. Taking into account this constraint seems to modify the picture drawing from flow equations only, in particular for the existence of non-Gaussian fixed points.

In this paper we will investigate the Ward constrained RG flow for a $\phi^6$-just renormalizable model in rank $4$. This model has been well defined and studied in the references \cite{BenGeloun:2011rc} and \cite{BenGeloun:2012yk} (see also \cite{Samary:2012bw} in the case of gauge invariant model). The motivation to be interested in such a model is the following: Note that the existence of a non-Gaussian fixed point can be understood from the following heuristic argument. As showed in \cite{Carrozza:2014rya}, the TGFT Feynman amplitudes can be analytically continued with respect to the group dimension $D$: $\U(1)\to \U(1)^D$. The power counting then indicate that there are two just-renormalizable models in rank 4: The $\phi^6$ melonic model in group dimension $D=1$ that we will describe in section \ref{section2} and which is asymptotically free; and the formal quartic melonic model in dimension $D=4/3$, which is also asymptotically free. Let $\vert\epsilon\vert\ll1$. The one-loop beta function for the quartic melonic model in dimension $D=4/3-\epsilon$ may be straightforwardly computed from the exact RG equation \eqref{Wetterich} we get:
\begin{align}\label{flowoneloop}
\beta^{(1)}_\lambda&=-3\epsilon \bar{\lambda}-2\eta^{(1)}\bar{\lambda}+4\bar{\lambda}^2\sqrt{\pi}+\mathcal{O}(\bar{\lambda}^2,\epsilon)\,,\\
\beta^{(1)}_{m^2}&=-\left(2+\eta^{(1)}\right)\bar{m}^2-8\bar{\lambda} \sqrt{\pi}+\mathcal{O}(\bar{\lambda}^2,\epsilon)\,,
\end{align}
where $\bar{\lambda}$ and $\bar{m}^2$ refer to the renormalized and dimensionless melonic coupling and mass parameter, and the one-loop anomalous dimension $\eta^{(1)}$ is given by:
\begin{equation}
\eta^{(1)}=4\bar{\lambda}\sqrt{\pi}\,.
\end{equation}
Equations \eqref{flowoneloop} admit a non trivial fixed point for the values:
\begin{equation}
\bar{\lambda}_*=-\frac{3\epsilon}{4\sqrt{\pi}}\,,\qquad \bar{m}^2_*=3\epsilon\,. \label{FP}
\end{equation}
Now, let us consider the Ward constraint. From the results on \cite{Lahoche:2019vzy}-\cite{Lahoche:2018ggd}, recalled in section \ref{section anomalous}, the Ward identities enforce a relation between beta functions and anomalous dimension. However, at one-loop order, this relation reduces to :
\begin{equation}
\beta^{(1)}_\lambda=-3\epsilon \bar{\lambda}-\eta^{(1)}\bar{\lambda}+\mathcal{O}(\bar{\lambda}^2,\epsilon)\,,
\end{equation}
which is nothing but the first equation \eqref{flowoneloop}. At one-loop, and at the first order in $\epsilon$, the non-Gaussian fixed point \eqref{FP} is then compatible with the Ward identities.
This fixed point has an attractive and a repulsive eigen-direction in the UV, as qualitatively illustrated on Figure \ref{fig0}; but it has the "wrong sign", a negative coupling which can break down the boundedness of the effective action. However, we have to keep in mind that $\epsilon \neq 0$ increase the strength of the inessential couplings; as a result, if such a formal fixed point survive in the limit $\epsilon\to 1/3$ when the $\phi^6$ couplings become marginal, it may be viewed as a UV attractive fixed point ensuring UV completion of the RG flow. The same phenomena have been pointed out in \cite{Carrozza:2014rya}. However, this paper aims to determine if such a fixed point remains compatible with Ward identities, by describing the flow at all orders in the coupling and the parameter $\epsilon$ and by using the EVE approach to solve the nonperturbative RG equation.
\\

\noindent
The outline of this paper is the following. In section \eqref{section2} we describe the model and argue in favour of a reduced family of $\phi^6$ couplings called \textit{non-branched}. In section \eqref{sec3} we use the EVE method to solve the nonperturbative RG equation. Finally, in section \eqref{sec4}, we briefly describe the RG flow in the vicinity of the Gaussian fixed point, allowing to recover the asymptotic freedom property. We then improve our analysis by implementing the so-called Ward constraint melonic flow and conclude that one new nontrivial fixed point can be found on this subspace.
\begin{center}
\includegraphics[scale=0.8]{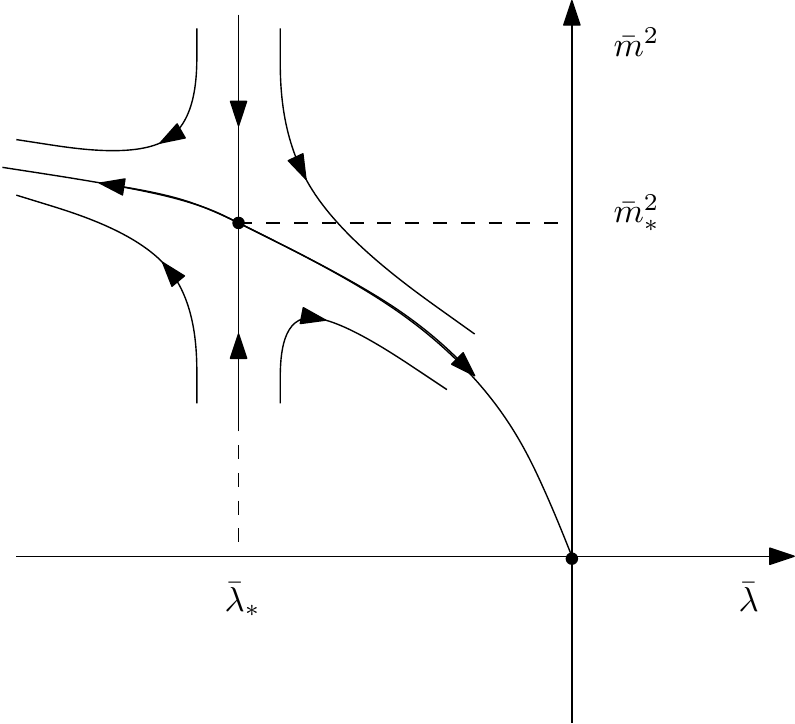}
\captionof{figure}{A qualitative description of the non-Gaussian fixed point in the $\epsilon$-expansion.}\label{fig0}
\end{center}


\section{Microscopic model and nonperturbtive RG}\label{section2}

As mentioned in the introduction, a group field theory (GFT) is a field theory whose fields are defined over $d$-copies of a group manifold $\mathrm{G}$. In this paper we focus on the $d$-dimensional torus and set $\mathrm{G}=\U(1)$, for a class of theories describing one complex field $\phi, \bar{\phi}:\U(1)^d\to \mathbb{C}$ with the free action:
\begin{equation}
S_{\text{kin}}[\phi, \bar{\phi}]=\int_{\mathrm{G}^{d}} d\textbf{\textbf{g}}\, \bar{\phi}({\textbf{g}} )\left(-\Delta_\textbf{g}+m^2\right)\phi({\textbf{g}} )\,,
\end{equation}
where $\textbf{g}:=(g_1,\cdots,g_d)\in \mathrm{G}^d$, and where $\Delta_\textbf{g}$ is the Laplace-Beletrami operator. For TGFTs, the interaction part is build as a sum of \textit{connected tensorial invariants} called bubbles. A bubble involves the same number of fields $\phi$ and $\bar{\phi}$, such that any variable of a field $\phi$ is contracted with the corresponding variable of a field $\bar{\phi}$, ensuring a proper unitary invariance per contracted indices. The elementary example is a mass-like term:
\begin{equation}
\int d\textbf{g} \,\bar{\phi}(g_1,g_2,\cdots,g_d )\phi(g_1,g_2,\cdots,g_d)\,.
\end{equation}
For higher order interactions, we may use of a convenient graphical notation. To each field $\phi$ (resp. $\bar{\phi}$) we associate a white (resp. black) node, with $d$ colored half edges hooked to him. Each of these colored edges corresponds to the group arguments, and the colors to their labels. For a given number of white dotes, the allowed interactions correspond to the number of different ways to hook the edges of the black and white nodes together, following their respective colors. As a result, each bubble may be graphically pictured as a $d$-colored bipartite regular graph, and the interaction part of the classical action is in full generality written as
\begin{equation}
S_{\text{int}}[\phi,\bar{\phi}]=g_1\,\vcenter{\hbox{\includegraphics[scale=0.8]{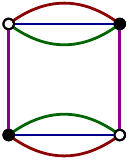} }}+\,g_2\,\vcenter{\hbox{\includegraphics[scale=0.8]{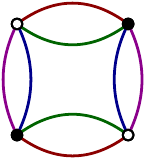} }}+\,g_3\,\vcenter{\hbox{\includegraphics[scale=0.8]{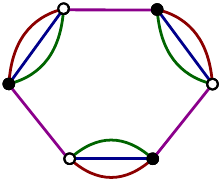}}}+\,g_4\,\vcenter{\hbox{\includegraphics[scale=0.8]{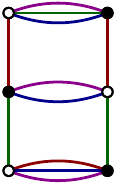}}}+\cdots
\end{equation}
It should be noted that, in the $\phi^6_4$ terminology, the number $4$ corresponds to the rank of the tensors $\phi$ and $\bar\phi$, i.e the dimension $d$. On the other hand the number $6$ is the maximal valence of the interactions. The spectrum of the Laplace-Beletrami operator introduces a canonical notion of scale, allowing to build a renormalization group for this class of theory. This work has been done in a series of different works since 2012 \cite{BenGeloun:2011rc}-\cite{Carrozza:2013wda}, investigating perturbative and nonperturbative aspects. The most difficulty with respect to standard quantum field theories or statistical models comes from the specific non-locality of the TGFT interactions. In order to deal with this specificity, a appropriate notion of locality has been addressed \cite{Carrozza:2012uv}:
\begin{definition}\label{locality}
For tensorial group field theories, connected bubbles are said to be local interactions.
\end{definition}
The Feynman graphs $\mathcal{G}$ indexing the amplitudes of the perturbation theory are $2$-simplex, i.e. a set of vertices, edges and faces : $\mathcal{G}:=(\mathcal{V}, \mathcal{L}\cup\mathcal{L}_{\text{ext}},\mathcal{F}\cup\mathcal{F}_{\text{ext}})$, respectively with cardinality $V$, $L$, $L_{\text{ext}}$, $F$ and $F_{\text{ext}}$. The sets of edges and faces split into internal and external subset, respectively without index and with index ‘‘$\text{ext}$''. For this let recall the definition of a faces:
\begin{definition}
A face is a maximal bi-colored subset of edges, including necessarily the color zero awarded from the Wick contractions. The subset can be closed and or opened respectively for close and open faces.
\end{definition}
For TGFTs as for any quantum field theory, perturbative amplitudes are indexed by Feynman graphs. The power counting has been established \cite{BenGeloun:2011rc}, and the superficial degree of divergence $\omega(\mathcal{G})$ corresponding to the graph $\mathcal{G}$ is :
\begin{equation}
\omega(\mathcal{G}):=-2L(\mathcal{G})+F(\mathcal{G})\,.\label{powercounting}
\end{equation}
The power counting allows to classify the theories following their renormalizability. For a given order in the perturbation expansion, the optimal graphs with respect to the power counting \eqref{powercounting} are called \textit{melonics}. They are the graphs having the greatest number of faces by fixing $L$; and one can show that for melonic diagrams the three numbers $V$, $L$ and $F$ are related as:
\begin{equation}
F=(d-1)(L-V+1)\,,
\end{equation}
Denoting as $\rho:=(d-1)(L-V+1)-F\geq 0$, the power counting may be rewritten as:
\begin{equation}
\omega(\mathcal{G})=\sum_k\left((d-3)k-(d-1)\right)v_k(\mathcal{G})+(d-1)-\frac{N(\mathcal{G})}{2}(d-3)-\rho(\mathcal{G}),
\end{equation}
where $v_k$ denotes the number of bubbles valence $2k$ (i.e. having $k$ whites nodes) and $N(\cG)=N$ in the number of external edges of the graph $\cG$. For melonic diagrams $\rho(\mathcal{G})=0$. The theory is said to be just-renormalizable if and only if $(d-3)k-(d-1)\leq 0$, the maximal valence being $k_0=(d-1)/(d-3)$. In particular, for $d=4$, we get $k_0=3$. The power counting suggests the existence of a just-renormalizable model involving $\phi^6$-interactions, with:
\begin{equation}
\omega(\mathcal{G})=-2v_1(\mathcal{G})-v_2(\mathcal{G})+3-\frac{N(\mathcal{G})}{2}-\rho(\mathcal{G})\,.\label{powercount}
\end{equation}
For $\rho=0$, only melonic diagrams with $N\leq 6$ are divergents. Moreover, one can easily show that $\rho\geq 1$, implying that for $N\geq 4$, all the superficially divergent diagrams are melonics see \cite{BenGeloun:2011rc}. An example of $2$-point superficially sub-leading divergent graph is given on Figure \ref{fig1}. From locality principle, definition \ref{locality}, the divergent subgraphs can be subtracted with local counter-terms. The procedure may be extended successfully to all orders of the perturbative expansion, allowing to prove a solid renormalizability theorems \cite{BenGeloun:2011rc}-\cite{Carrozza:2013wda}.
\begin{center}
\includegraphics[scale=1]{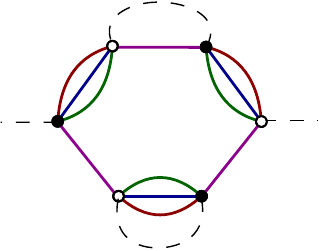}
\captionof{figure}{A sub-divergent $2$-point diagram builds from a $6$-point vertex. The dotted edges correspond to the Wick-contractions. From a direct computation $\omega=-2\times 2+(3+1)=0$. }\label{fig1}
\end{center}
\noindent
Let $\cG$ be a melonic diagram, then its vertex bubbles have to be melonic. We recall briefly the definition of the melonic bubbles:
\begin{definition}
Any melonic bubble $b_\ell$ of valence $\ell$ may be deduced from the elementary melon $b_1$:
\begin{equation}
\vcenter{\hbox{\includegraphics[scale=1]{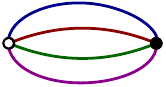} }}\,,
\end{equation}
replacing successively $\ell-1$ colored edges as follows, defining the insertion operator $\mathfrak{R}_{i}$:
\begin{equation}
\vcenter{\hbox{\includegraphics[scale=1]{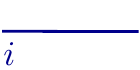} }}\underset{\mathfrak{R}_{i}}{\longrightarrow}\vcenter{\hbox{\includegraphics[scale=1]{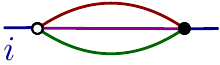} }}\,.
\end{equation}
Then $b_\ell := \left(\prod_{\alpha=1}^{\ell-1}\mathfrak{R}_{i_\alpha}\right) b_1$.
\end{definition}
The melons graphs were introduced in the colored random tensors models, where they arise as leading order graphs in the $1/N$-expansion\footnote{ Remark that in the terminology {\it ``$1/N$-expansion''}, $N$ refer to the size of the tensors but is not the number of external edges of the graph $\cG$ which is also denoted by $N$.}. This expansion is characterized with an exponent $\varpi$ playing the same role as the genus for random matrices, the \textit{Gurau degree}, vanishing for melons. The Abelian $\phi^6_4$-melonic model \footnote{The expression {\it ``melonic''} referred to the leading order contribution i.e. the partition and correlation functions admit perturbative expansions which are dominated by peculiar triangulations of spheres called melons.}, defined with the classical action:
\begin{equation}
S[\phi,\bar{\phi}]=S_{\text{kin}}[\phi,\bar{\phi}]+\lambda_{4}\,\sum_{i=1}^4 \vcenter{\hbox{\includegraphics[scale=0.8]{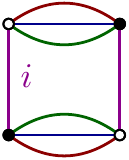} }}+\lambda_{6,1}\,\sum_{i=1}^4\vcenter{\hbox{\includegraphics[scale=0.8]{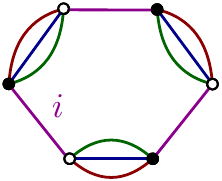} }}+\lambda_{6,2}\,\sum_{i<j}\vcenter{\hbox{\includegraphics[scale=0.8]{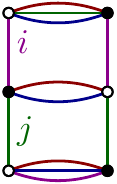} }}\,,\label{classicAction1}
\end{equation}
has been showed to be just-renormalizable, and a BPHZ theorem has been also proved \cite{BenGeloun:2011rc}.

\noindent
The renormalization group flow describes the change of the couplings in the effective action when UV degrees of freedom are integrated out. The functional renormalization group (FRG) formalism is a specific way to build such an evolution in the theory space of all possible actions. It has been showed to be a promising theoretical framework for tensorial field theory, allowing to deal with the specific non-locality of the interactions and to investigate nonperturbative regime. In the FRG formalism, the quantum model is described by a one-parameter family of partition functions $\{\mathcal{Z}_k\}$, $k\in\mathbb{R}$, from a microscopic action $S$ in the deep UV ($k=\Lambda$) to an effective action $\Gamma$ in the deep IR ($k=0$). For a just-renormalizable models, the microscopic scale $\Lambda$ becomes irrelevant, and may be removed in the continuum limit $\Lambda\to\infty$. If $k$ walks around the interval $[\Lambda,0]$, we run through successive effective models, with effective average action $\Gamma_k$. The step partition function $\mathcal{Z}_k(J,\bar{J})$ is defined as:
\begin{equation}
\mathcal{Z}_k(J,\bar{J}):=\int d\phi d\bar{\phi} \,e^{-S[\phi,\bar{\phi}]-R_k[\phi,\bar{\phi}]+\bar{J}\cdot \phi+\bar{\phi}\cdot J}\,,\quad \forall k\in [0,\Lambda[\,. \label{Z}
\end{equation}
where $a\cdot b:= \int d\textbf{g} \,a(\textbf{g})\,b(\textbf{g})$. The driving or \textit{coarse-graining} along the RG flow is ensured by the regulator function $R_k[\phi,\bar{\phi}]$, which behaves like a momentum dependent mass term,
\begin{equation}
R_k[\phi,\bar{\phi}]:=\int d\textbf{g}\,\bar{\phi}(\textbf{g})\,r_k(-\Delta){\phi}(\textbf{g})\,,
\end{equation}
ensuring that the UV degrees of freedom with respect to the running scale $k$ are integrated out, with boundary conditions:
\begin{equation}
\Gamma_{k=\Lambda}=S\,,\qquad \Gamma_{k=0}=\Gamma\,.\label{boundary}
\end{equation}
The regulator $r_k(-\Delta)$ depends on the spectrum values of the Laplace-Beletrami operator, and is positive defined. For convenience, we work in the momentum representation, the Fourier components of the fields $\phi$ and $\bar{\phi}$ becoming discrete tensors on $\mathbb{Z}^4$:
\begin{equation}
\phi(g_1,g_2,g_3,g_4)=\sum_{\vec{p}\in\mathbb{Z}^4} T_{p_1,p_2,p_3,p_4}e^{i\sum_j p_j\theta_j} \,,\quad \bar{\phi}(g_1,g_2,g_3,g_4)=\sum_{\vec{p}\in\mathbb{Z}^4} \bar{T}_{p_1,p_2,p_3,p_4}e^{-i\sum_j p_j\theta_j} \,,
\end{equation}
the $\theta$--coordinates being angle variables $\theta_j\in[0,2\pi[$, such that $g_j=e^{i\theta_j}$. In Fourier representation, $r_k$ becomes a function of the square of the momentum only, having generically the following structure:
\begin{equation}
r_k(\vec{p}\,^2):=k^2\,f(\vec{p}\,^2/k^2)\,.
\end{equation}
The function $f$ have to be positive defined as well, and have to satisfy some requirements with respect to the boundary conditions \eqref{boundary}, among which:
\begin{enumerate}
\item $\lim_{k\to\Lambda}r_k(\vec{p}\,^2)\gg 1\,,$
\item $\lim_{k\to0}r_k(\vec{p}\,^2) =0\,,$
\item $r_k(\vec{p}\,^2>k^2)\simeq 0\,.$
\end{enumerate}
The effective average action $\Gamma_k$ is defined as a slightly modified Legendre transform of the free energy $W_k:=\ln \mathcal{Z}_k$:
\begin{equation}
\Gamma_k[M,\bar{M}]+R_k[M,\bar{M}]=\bar{J}\cdot M+\bar{M}\cdot J-W_k[J,\bar{J}]\,,\label{average}
\end{equation}
where the means fields $M$ and $\bar{M}$ are themselves tensor fields, defined as:
\begin{equation}
M_{\vec{p}}:= \frac{\partial W_k}{\partial \bar{J}_{\vec{p}}}\,,\qquad \bar{M}_{\vec{p}}:= \frac{\partial W_k}{\partial {J}_{\vec{p}}}\,.\label{M}
\end{equation}
The presence of the $R_k$-term in the definition \eqref{average} enforce the initial condition $\Gamma_{k=\Lambda}=S$. Moreover, it means that the effective $2$-point function $G_k$ have to be related with the second derivative of the effective action i.e. $\Gamma^{(2)}_k:=\partial_M\partial_{\bar{M}}\Gamma_k$ as:
\begin{equation}
G_k(\vec{p},\vec{p}\,^\prime):=\left(\Gamma^{(2)}_k+r_k\mathbb{I}\right)^{-1}(\vec{p},\vec{p}\,^\prime)\,,
\end{equation}
where $\mathbb{I}$ designates the identity matrix. The effective average action move through the theory space with the running scale $k$, and its evolution obeys to the exact flow equations known as \textit{Wetterich-Morris equation} :
\begin{equation}
\dot{\Gamma}_k=\sum_{\vec{p}\in\mathbb{Z}^4} \dot{r}_k(\vec{p}\,^2)\left(\Gamma^{(2)}_k+r_k\mathbb{I}\right)^{-1}(\vec{p},\vec{p}\,)\,,\label{Wetterich}
\end{equation}
where the dot designates the derivative with respect to the normal coordinate along the flow curves: $s:=\ln k$. Despite its simplicity, this equation is very difficult to solve exactly; and requires appropriate approximation schemes. This difficulty may be tracked by taking in the Wetterich equation the successive derivations with respect to the means fields $M$ and $\bar{M}$, generating an infinite hierarchical system of coupled equations. For TGFTs, the non-locality of the interactions introduce a substantial difficulty, discarding some powerful tools used for standard quantum fields. Until a recent year, only \textit{truncation method}, which stop crudely the hierarchical system, seemed to be relevant to solve the equation \eqref{Wetterich}. Some progress in solving the flow equation beyond truncation method have been made in a series of recent works \cite{Lahoche:2019vzy}-\cite{Lahoche:2018vun}. The method, called \textit{effective vertex expansion} (EVE) allows to truncate ‘‘smoothly" the infinite hierarchical system, closing it around marginal operators. The strategy is to use relevant and marginal operators to drive the flow of highest order effective vertices, all expressed in terms of a reduced set of effective functions. In this approach, moreover, a fixed point for the reduced system has to be as well a global fixed point. Until now, there are four limitations for this methods. First of all, it work well only in the UV sector $\Lambda\gg k\gg 1$, where only leading order contributions survive. This condition play an important role, the properties of the leading sector being used to close the hierarchical system. Secondly, it seems difficult to use it for branching interactions like the interaction associated to the coupling $\lambda_{6,2}$ in equation \eqref{classicAction1}. Thirdly, the method is difficult to extend for highest order of the derivative expansion. Finally, the EVE method has not been applied beyond the \textit{symmetric phase}, where the means field vanish. Among the relevant properties of the region of the phase space reached from an expansion around $M=\bar{M}=0$, we mention the following, discussed in \cite{Lahoche:2018hou}-\cite{Lahoche:2018oeo}:
\begin{property}\label{property1}
In the symmetric phase, all the odd effective vertex functions, having not the same number of derivatives with respect to $M$ and $\bar{M}$ vanish. Moreover, the effective $2$-point function $G_k(\vec{p},\vec{p}\,^\prime)$ is diagonal:
\begin{equation}
G_k(\vec{p},\vec{p}\,^\prime):=G_k(\vec{p}\,)\delta(\vec{p},\vec{p}\,^\prime)\,.
\end{equation}
\end{property}
To fix the initial conditions, we restrict our attention on the non-branching sector, defined as follows:
\begin{definition}
A non branching melonic bubble of valence $\ell$, $b_\ell^{(i)}$ is labeled with a single index $i\in\llbracket 1,4\rrbracket$, and defined such that:
\begin{equation}
b_\ell^{(i)}:= \left(\mathfrak{R}_{i}\right)^{\ell-1}\,b_1\,.
\end{equation}

\end{definition}
Figure \ref{fig2} provides the generic structure of melonic non-branching bubbles.

\begin{center}
\begin{equation*}
\vcenter{\hbox{\includegraphics[scale=0.9]{Melon1.pdf} }} \,\underset{\mathfrak{R}_{i}}{\longrightarrow}\, \vcenter{\hbox{\includegraphics[scale=0.9]{Melon2.pdf} }}\,\cdots \underset{\mathfrak{R}_{i}}{\longrightarrow}\, \vcenter{\hbox{\includegraphics[scale=0.8]{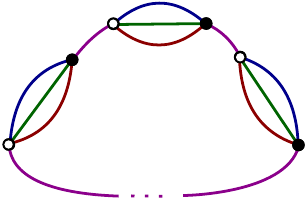} }}\underset{\mathfrak{R}_{i}}{\longrightarrow}\cdots
\end{equation*}
\captionof{figure}{Structure of the non-branching melons, from the smallest one $b_2$.} \label{fig2}
\end{center}
As mentioned before, the method we consider in this paper to deduce the flow equations work well only for the non-branching sector. The same point has been discussed for another family of graphs called \textit{pseudo melons} in \cite{Lahoche:2018oeo}; and it is the first source of limitation for our incoming conclusions. The interest of this sector in the fixed point investigation have been pointed out in \cite{Carrozza:2016tih} for a non-Abelian TGFT over $\SU(2)^3$ with closure constraint. Because we expect that the UV behaviour of this model and the model defined by the action \eqref{classicAction1} are the same, the restriction to the non-branching sector seems to be not so bad as a starting condition. There is another interest for this restriction, justifying the terminology "sector''. In the UV domain $\Lambda \gg k\gg 1$, the corresponding reduced theory space is stable under the renormalization group transformations. Indeed, deriving the flow equation \eqref{Wetterich} with respect to $M$ and $\bar{M}$, and from the property \ref{prop1}, we get that the flow equation for $\Gamma^{(n)}_k$ involves $\Gamma^{(n+2)}_k$, $\Gamma^{(n)}_k$, and smallest effective vertices. All these effective vertices can be labelled with a bubble drawing the pattern following which external momenta are pairwise identified, and corresponding to the \textit{boundary graph} of the relevant Feynman graphs involved in the perturbative expansion of these effective vertex functions. From a direct inspection, it can be easily checked recursively that, starting with effective vertices indexed with non-branching melonic bubbles as a building block, we do not generate leading order one-loop contributions outside of the non-branching subspace. The calculations of the section \ref{sectionflow} support explicitly this argument. \\

\noindent
From these considerations, the microscopic action $S$, fixed for some fundamental UV scale $\Lambda$ is the following:
\begin{equation}
S[T,\bar{T}]=\sum_{\vec{p}\in\mathbb{Z}^\droit} \, \bar{T}_{\vec{p}}\left(Z_{-\infty}\vec{p}\,^2+Z_2m^2_0\right)T_{\vec{p}}+Z_4\lambda_{4}\,\sum_{i=1}^4 \vcenter{\hbox{\includegraphics[scale=0.8]{Melon12.pdf} }}+Z_6\lambda_{6}\,\sum_{i=1}^4\vcenter{\hbox{\includegraphics[scale=0.8]{Melon21.pdf} }}\,,\label{classicAction2}
\end{equation}
where $Z_{-\infty}$, $Z_2$, $Z_4$ and $Z_6$ denotes respectively the wave function, mass and couplings counter-terms. From renormalizability theorem, they allows to cancel all the divergences occurring in the perturbative expansion. In standard nonperturbative RG analysis, these counter-terms are not explicitly introduced, and the choice of the initial conditions is not extensively discussed. The reason why we discuss them has been explained in a recent work \cite{Lahoche:2018vun} for a melonic $\phi^4$ model. The non perturbative equation \eqref{Wetterich} is divergence free due to the regulator function $\dot{r}_s$. However, we will discuss the compatibility between flow equations and Ward identities; and some UV divergences occurs in them, as a consequence of the non-locality of the interactions. Counter-terms are then essentials to deal with these specific divergences. The index “$-\infty$" refers to the fact that the finite part of the counter-terms in fixed in the deep infrared limit $k\to 0$ or $s\to-\infty$. Note that, the limit may be formal, and simply means that it is chosen for very small $k$, i.e. at a scale so far from the domain $k\ll 1$. \\

\noindent
To conclude this section, we provide some central notion for our analysis called the \textit{canonical dimension}. In standard quantum field theories (QFTs), the dimension of the interactions is closely related to their renormalizability. Interactions with positive or vanishing (momentum) dimensions are renormalizable, while the ones with negative dimensions are non-renormalizable. For GFTs, however, the situation is more subtle. There is no reference scale in the classical action \eqref{classicAction2}, and the sums over $\mathbb{Z}^d$ are dimensionless in contrast with integrations over space-time in ordinary QFT. There are two ways to introduce a dimension in the GFT framework and recover the standard classification following the dimension of the couplings. The first one is to make contact with physical quantities. In particular, for non-Abelian models over $\SU(2)$, the spectrum of the Laplacian may be related with the size of the area operator in loop quantum gravity (LQG), the result can be extended for Abelian cases. The second strategy is to fix the dimension from the renormalization group flow. Indeed, the leading order scaling of the operators concerning the UV cut-off may be viewed as a dimension; and from the definition of a just-renormalizable model, one expects that just-renormalizable interactions scale logarithmically with the cut-off, and then have zero dimension. For melonic diagrams in the UV, one can easily check the canonical dimension of the operators from the power-counting law \eqref{powercounting}. Any leading order $N$-points functions build only with $6$-points interactions scale as:
\begin{equation}
\omega=3-\frac{N}{2}\,. \label{dim}
\end{equation}
In particular, the proper scaling of the leading order $6$-point functions vanish, meaning that it scales logarithmically with the UV cut-off. Following the definition of the canonical dimension as the leading order scaling of the quantum corrections, it makes sense to associate a dimension zero for the coupling $\lambda_6$. For the same reason, setting $N=4$, the dimensions of the coupling $\lambda_4$ has to be $1$. Indeed, from the definition of renormalizability, we expect a proliferation of just-renormalizable interactions, which have to fix the leading scaling. Note that this condition is compatible with the ‘‘cost" $-1$ of the $\phi^4$ melonic bubbles in the power counting \eqref{powercounting}. Finally, the dimension of the mass parameter has to be $2$. For interactions with valence $N/2$, the canonical dimension is explicitly given from formula \eqref{dim}. We then recover the standard classification, renormalizable interactions having a positive dimension, and non-renormalizable ones having negative dimension. For an extended discussion, the reader may consult \cite{Lahoche:2015ola}.

\section{Solving RG equation in the non-branching sector}\label{sec3}

In this section, we solve the exact flow equation \eqref{Wetterich} using the EVE approximation scheme introduced in \cite{Lahoche:2019vzy}-\cite{Lahoche:2018vun}. The melonic model that we consider being very close to the non-branching pseudo-melonic sector discussed in \cite{Lahoche:2018ggd}, we give only the main steps of the proof. As explained in the previous section, the strategy is to close the hierarchical system of coupled equations deduced from \eqref{Wetterich}, that is, to explain the effective vertex $\Gamma^{(8)}_k$ in terms of $\Gamma^{(6)}_k$, $\Gamma^{(4)}_k$ and $\Gamma^{(2)}_k$; using the essential and marginal couplings to drag the RG flow of highest effective vertices. Moreover, the method go beyond a ‘‘smooth" truncation, in the sense that EVE allows capturing the momentum dependence of the effective vertex, which play a role of the anomalous dimension. To be more precise, the Ward identity allows computing the momentum derivative of the effective $4$-point vertices in term of the marginal and essential couplings, improving the crude truncation which does not consider this additional contribution. Indeed, as pointed out in \cite{Lahoche:2019vzy}, took into account the momentum dependence of the $4$-point vertex leads to discard a line of singularity and to extend maximally the symmetric phase region. We start from the derivation of the hierarchical equations in the non-branching melonic sector up to $6$-point effective local vertices. In a second time, we build the structure equation for the $\Gamma^{(8)}_k$ vertex and use Ward identities to compute the derivative of the effective $4$-point vertex. As a result, we get a set of four autonomous coupled equations, involving only relevant and marginal dimensionless couplings. Finally, we discuss an additional constraint coming from Ward identity linking $\Gamma^{(4)}_k$ and $\Gamma^{(2)}_k$. As pointed out in \cite{Lahoche:2019vzy}, in the deep UV, this equation may be turned locally in the flow as a constraint linking beta functions for $4$ and $2$-point relevant and marginal couplings. This additional constraint has to be solved simultaneously with the flow equations, reducing the dimension of the phase space, from three to two.

\subsection{RG equations for marginal local couplings}\label{sectionflow}
Deriving a first time with respect to $M$ and $\bar{M}$ the exact flow equation \eqref{Wetterich}, we get an equation for $\dot{\Gamma}^{(2)}$, involving $\Gamma^{(4)}$ and $\Gamma^{(2)}$:
\begin{equation}
\dot{\Gamma}_k(\vec{p}\,)=-\sum_{\vec{q}\in\mathbb{Z}^d} \Gamma^{(4)}_{k}(\vec{p},\vec{p},\vec{q},\vec{q}\,) \dot{r}_k\,G_k^2(\vec{q}\,)\,,\label{eq1}
\end{equation}
where we used property \ref{property1}. As explained in the previous section, the effective vertex function may be indexed with a non-branching bubble corresponding to the boundary of the graphs indexed the amplitudes in their perturbative expansion. For non-branching bubbles, these boundaries are indexed with a single index. Then as a consequence, the non-branching effective vertices $\Gamma^{(n)}$, for $n>2$ decompose as:
\begin{equation}
\Gamma^{(2n)}_k=\sum_{i=1}^\droit \,\Gamma^{(b_n^{(i)})}_k\,.\label{decomp}
\end{equation}
The structure of the partial $n$-point functions $\Gamma^{(b_n^{(i)})}_k$ have been extensively discussed in \cite{Lahoche:2019vzy}. The boundary graph dictates the ways to identify external momenta. Formally, it corresponds to a product of Kronecker deltas following the path drawing by the corresponding bubble interaction. To simplify the notations, we represent these products of deltas with $d$-colored regular bipartite graphs as well, indexed with external momenta. For instance:
\begin{align}
\vcenter{\hbox{\includegraphics[scale=0.9]{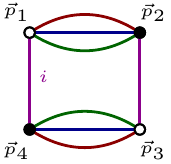} }}&=\delta_{p_{1i}p_{4i}}\delta_{p_{2i}p_{3i}}\,\prod_{j\neq i}\delta_{p_{1j}p_{2j}}\delta_{p_{3j}p_{4j}}\,,\\
\vcenter{\hbox{\includegraphics[scale=0.8]{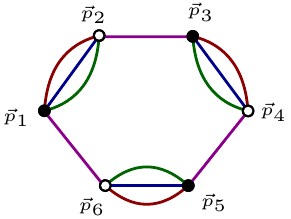} }}&=\delta_{p_{1i}p_{6i}}\delta_{p_{2i}p_{3i}}\delta_{p_{4i}p_{5i}}\prod_{j\neq i}\delta_{p_{1j}p_{2j}}\delta_{p_{3j}p_{4j}}\delta_{p_{5j}p_{6j}}\,.
\end{align}
The partial effective vertex functions $\Gamma^{(b_n^{(i)})}_k$ then may be written as:
\begin{equation}
\Gamma^{(b_n^{(i)})}_k(\{\vec{p}_\ell\})=:\pi_k^{(b_n^{(i)})}(p_{\ell\,i})\times\sym\left(\vcenter{\hbox{\includegraphics[scale=0.9]{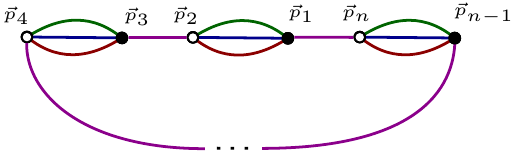} }}\right)\,,
\end{equation}
where $\sym$ denotes the permutation of the external momenta. For instance:
\begin{equation}
\sym \left(\vcenter{\hbox{\includegraphics[scale=0.9]{Melon13.pdf} }}\right)=2\left(\vcenter{\hbox{\includegraphics[scale=0.9]{Melon13.pdf} }}+\vcenter{\hbox{\includegraphics[scale=0.9]{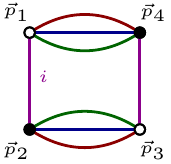} }}\right)\,,
\end{equation}
the factor $2$ coming from the fact that permuting both the black and white nodes do not change the configuration. The kernels $\pi_k^{(b_n^{(i)})}:\mathbb{Z}^n\to\mathbb{R}$ depend on the $i$-th component of the external momenta. The zero-momenta values of the effective vertex functions are related with the effective couplings constants at scale $k$. For relevant and marginal couplings, we set:
\begin{equation}
\Gamma^{(2)}_k(\vec{0}\,)=:m^2(k)\,,\quad \Gamma^{(4)}_k(\{\vec{0}\})=:(2!)^2\lambda_4(k)\,\quad \Gamma^{(6)}_k(\{\vec{0}\})=: (3!)^2\lambda_6(k)\,.\label{rencond}
\end{equation}
the $k$ dependence on the couplings allowing to differentiate them from the ‘‘bare" couplings. From definition of the kernels $\pi_k^{(b_n^{(i)})}$, we have therefore : $\pi_k^{(b_2^{(i)})}(0)=\lambda_4(k)$ and $\pi_k^{(b_3^{(i)})}(\{0\})=\lambda_6(k)$. For the rest of this section, we left the arguments for zero-momenta functions when it is unambiguous. Inserting the decomposition \eqref{decomp} in the flow equation \eqref{eq1}, we get, graphically:
\begin{equation}
\dot{\Gamma}_k(\vec{p}\,)=-2\sum_{i=1}^d \pi_k^{(b_2^{(i)})}(p_i) \left(\vcenter{\hbox{\includegraphics[scale=0.9]{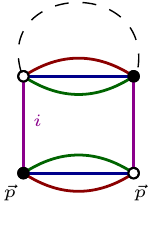} }}+\,\vcenter{\hbox{\includegraphics[scale=0.9]{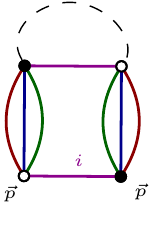} }}\right)\approx-2\sum_{i=1}^d \pi_k^{(b_2^{(i)})}(p_i) \,\left(\vcenter{\hbox{\includegraphics[scale=0.9]{Melon13int.pdf} }}\right)\,, \label{eq2}
\end{equation}
where the dashed line corresponds to the contraction with the propagator $\dot{r}_kG_k^2$. For the last term we only kept the melonic contractions, creating three internal faces, the second contribution having a relative scaling $k^{-2}$ with respect to the melonic one, and may be discarded in the UV sector $k\gg 1$. The rule follows as soon as we remains in this domain; and for higher order interactions we will keep only the melonic contraction.
We get, for $\Gamma^{(4)}_k$ and $\Gamma^{(6)}_k$, taking successive derivatives of the equation \eqref{Wetterich}, and vanishing their external momenta, we get from renormalization conditions \eqref{rencond}:
\begin{equation}
(2!)^2 \dot{\lambda}_4=-12\pi_k^{(b_3^{(1)})}\,\left(\vcenter{\hbox{\includegraphics[scale=0.9]{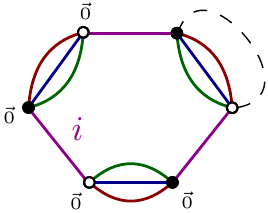} }}\right)+4\times (2!)^2 \big(\pi_k^{(b_2^{(1)})}\big)^2 \left(\vcenter{\hbox{\includegraphics[scale=0.9]{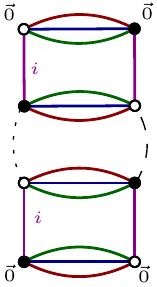} }}\right)\,,\label{eq3}
\end{equation}
where the dotted line on the last diagram corresponds to the contraction with the effective propagator $G_k$. All the numerical coefficients count the number of independent melonic contractions, or arises from the derivation of the equation \eqref{Wetterich}. This is the case of the factor $4$, which is in fact $2+2$: a first factor $2$ coming from the derivation of $G_k^2$ in the equation \eqref{eq1}, a second factor $2$ coming from a odd term, discarded in the calculation of equation \eqref{eq1}. In the same way:
\begin{align}
\nonumber(3!)^2\dot{\lambda}_6=&-4\times (3!)^2 \pi_k^{(b_4^{(1)})}\left(\vcenter{\hbox{\includegraphics[scale=0.9]{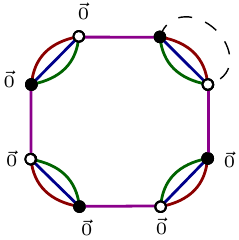} }}\right)+2(3!)^3 \pi_k^{(b_3^{(1)})}\pi_k^{(b_2^{(1)})}\left(\vcenter{\hbox{\includegraphics[scale=0.9]{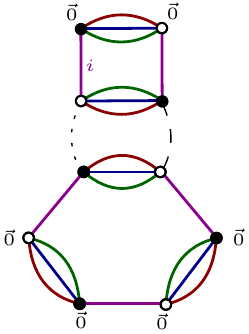} }}\right)\\
&-2^3(3!)^2\big( \pi_k^{(b_2^{(1)})}\big)^3\left(\vcenter{\hbox{\includegraphics[scale=0.9]{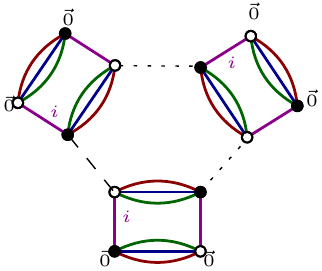} }}\right)\,.\label{eq4}
\end{align}
The mass flow may be fixed from the renormalization conditions \eqref{rencond}, setting $\vec{p}=\vec{0}$ on both sides of equation \eqref{eq1}. All the diagrams may be easily computed. They involve only a single loop, creating $3$ internal faces. Denoting as $p$ the external momentum running through the external face of color $i$, the explicit expression for the loop of length $j$, $L_j(p)$ writes as:
\begin{equation}
L_j(p):=\sum_{\vec{q}\in\mathbb{Z}^4} \delta_{q_1p} \,\dot{r}_k\,G_k^{j+1}(\vec{q}\,)\,.\label{sum}
\end{equation}
Therefore, the equations \eqref{eq2}, \eqref{eq3} and \eqref{eq4} may be written explicitly as follow:
\begin{align}\label{system1}
\dot{m}^2&=-2\lambda_4 L_1(0)\,,\\
\dot{\lambda}_4&=-3\lambda_6 L_1(0)+4\lambda_4^2 \,L_2(0)\,,\\
\dot{\lambda}_6&=-4 \pi_k^{(b_4^{(1)})} L_1(0)+12 \lambda_4\lambda_6 L_2(0)-8\lambda_4^3L_3(0)\,,
\end{align}
where we denote by $\dot x:=k\frac{dx}{dk}$. To close the system, we have to express the remaining piece $\pi_k^{(b_4^{(1)})}$, the effective coupling with valence $4$, in terms of the relevant and marginal couplings. This aims imply to understand independently the dependence of the $4$, $6$ and $8$-points effective vertices with respect to the effective marginal and essential couplings at scale $k$. Investigating the structure of the leading perturbative Feynman graphs building the effective vertices, \textit{assumed to be analytic functions of the effective couplings}, and from renormalization conditions defining effective couplings at scale $k$, we deduce some relations between effective vertices and renormalizable couplings. The assumption about analyticity of the effective vertex functions ensures the validity of these relations out of the perturbative domain, as long as we remain into the symmetric phase. Indeed, expanding around non-zero vacuum break the diagonal condition given by property \ref{property1}, which is satisfied in the perturbative sector as well. \\

\subsection{Closing hierarchy: The EVE method}

\noindent
In a first time we restrict our attention on the quartic sector, whose graphs building the effective functions are made with quartic melonic interactions only. We have the following proposition:
\begin{proposition}\label{prop4}
Let us denote as $\bar{\pi}_k^{(b_n^{(i)})}$ the leading order effective vertices of valence $n$ build of quartic melonic interactions only. For $n=2,3$ and $4$, their expressions in terms of the bare coupling $Z_4\lambda_4$ are the following:
\begin{align}
\bar{\pi}_k^{(b_2^{(i)})}&=\frac{Z_4\lambda_4}{1+2Z_4\lambda_4\,\mathcal{A}_{k,2}}\,, \label{prop1}\\
\bar{\pi}_k^{(b_3^{(i)})}&=\frac{8}{3}\big(\bar{\pi}_k^{(b_2^{(i)})}\big)^3 \,\mathcal{A}_{k,3}\,,\label{prop2}\\
\bar{\pi}_k^{(b_4^{(i)})}&=-2\big(\bar{\pi}_k^{(b_2^{(i)})}\big)^4 \,\mathcal{A}_{k,4}+\frac{16}{3}\big(\bar{\pi}_k^{(b_2^{(i)})}\big)^5\,\big(\mathcal{A}_{k,3}\big)^2\label{prop3}\,,
\end{align}
where $\mathcal{A}_{k,n}\equiv \mathcal{A}_{k,n}(0)$, with
\begin{equation}
\mathcal{A}_{k,n}(p):=\sum_{\vec{q}\in\mathbb{Z}^4} \,\delta_{pq_1}\,G_k^{n}(\vec{q}\,)\,.
\end{equation}
\end{proposition}
To prove this proposition, we recall the following lemma, already considered in \cite{Lahoche:2018oeo}:
\begin{lemma}
Let $\mathcal{G}$ be a 1PI $2N$-points diagram with more than one vertex. It has $N$ boundary vertices to which external edges are hooked, $d-1$ external faces per external vertex, and $N$ external faces of the same color running through the interior of the diagram. \label{propmelonfaces}
\end{lemma}
\noindent
\textit{Proof:} We will proceed recursively for each $n$, and we will indicates only the first steps of the proof. More details may be found in \cite{Lahoche:2018oeo}. \\

\noindent
$\bullet$ \textit{4-point effective vertices.} Let us denote as $Z_4\lambda_4\,\Pi_4$ the part of $\bar{\pi}_k^{(b_2^{(i)})}$ made of at least two vertices. We have, setting $i=1$:
\begin{equation}
\bar{\pi}_k^{(b_2^{(1)})}=Z_4\lambda_4\,(1+\Pi_4)\,.\label{eqstart1}
\end{equation}
Let us consider the structure of $\Pi_4$. Taking into account the face-connectivity and the lemma \ref{propmelonfaces}, the boundary vertices may be such that the two internal faces of the same color running on the interior of the diagrams building $\Pi_4$ pass through of them. As a result, we expect the following structure, compatible with the boundary graph $b_2^{(1)}$:
\begin{equation}
-Z_4\lambda_4\,\Pi_4=\vcenter{\hbox{\includegraphics[scale=0.6]{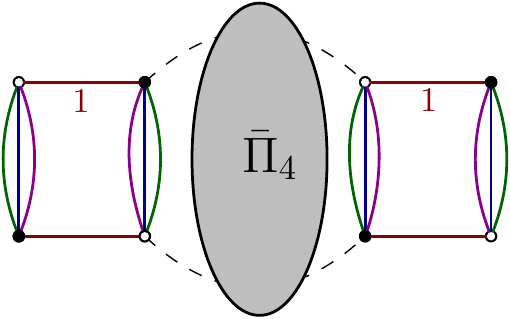} }}\,,
\end{equation}
where the grey disk indexed with $\bar{\Pi}_4$ is itself a sum of Feynman graphs. It is easy to verify that other configurations of the boundary vertices move away from the leading order. Extracting the one-particle irreducible part $\bar{\Pi}_4^\prime$ of $\bar{\Pi}_4$, we get, graphically as well:
\begin{equation}
\vcenter{\hbox{\includegraphics[scale=0.5]{pin.pdf} }}=\vcenter{\hbox{\includegraphics[scale=0.5]{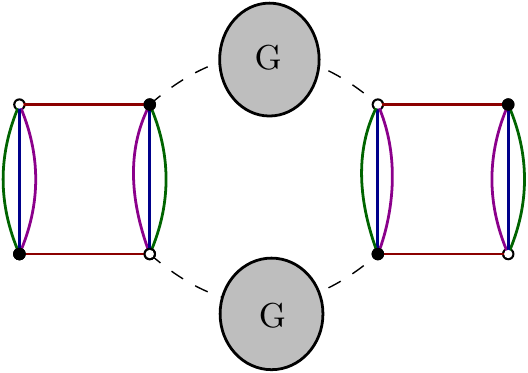} }}+\, \vcenter{\hbox{\includegraphics[scale=0.5]{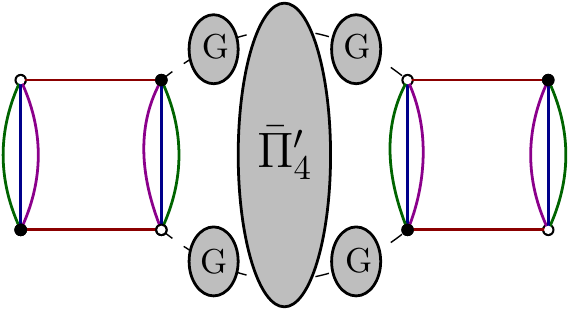} }}
\end{equation}
where the grey disks indexed by $G$ denote the leading order effective propagator\textit{ in the quartic sector} (we left the index $k$ to simplify the figures). The remaining contribution $\Pi_4^\prime$ is at least of order $1$, that is, build with a single vertex. Isolating this vertex, the next order terms build an effective vertex made of 1PI graphs having at least two vertices, corresponding to that we called $\Pi_4$. As a result, it is easy to check the following closed equation:
\begin{equation}
\vcenter{\hbox{\includegraphics[scale=0.5]{pin2.pdf} }}=\vcenter{\hbox{\includegraphics[scale=0.5]{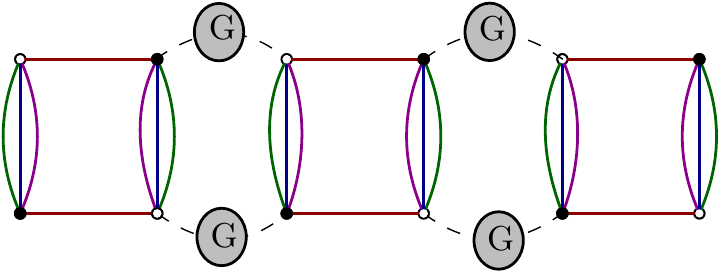} }}+\,\vcenter{\hbox{\includegraphics[scale=0.5]{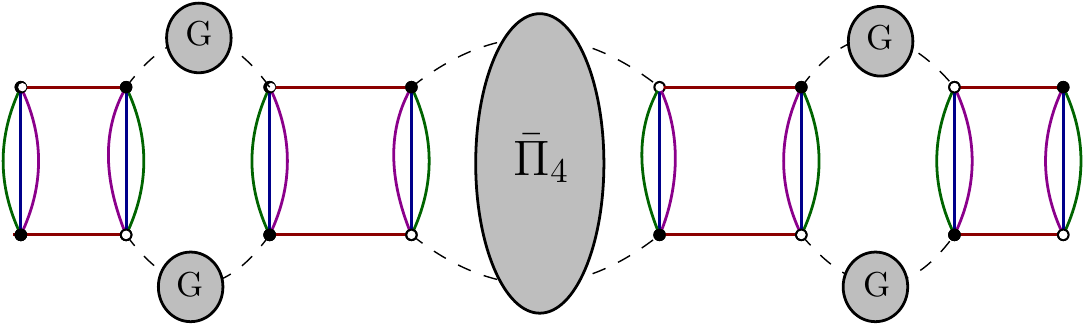} }}\,,
\end{equation}
which can be formally solved recursively as:
\begin{equation}
-Z_4\lambda_4\,\Pi_4= \vcenter{\hbox{\includegraphics[scale=0.5]{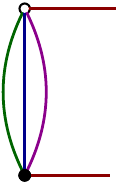} }} \left\{ \sum_{n=1}^\infty\left(\vcenter{\hbox{\includegraphics[scale=0.5]{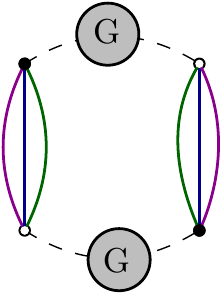} }}\right)^n\right\} \vcenter{\hbox{\includegraphics[scale=0.5]{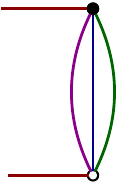} }}\,.
\end{equation}
Consider now equation \eqref{eqstart1}, we then get:
\begin{equation}
\bar{\pi}_k^{(b_2^{(1)})}=Z_4\lambda_4\left(1-\vcenter{\hbox{\includegraphics[scale=0.5]{pimiddle2.pdf}}}\right)^{-1}\,,\label{step1}
\end{equation}
where we used the following graphical conventions,
\begin{equation}
\vcenter{\hbox{\includegraphics[scale=0.5]{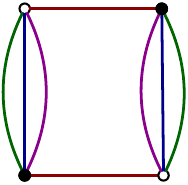} }} \equiv -Z_4\lambda_4\,.
\end{equation}
Finally, the effective one-loop diagram on the denominator of \eqref{step1} may be easily computed recursively, leading to:
\begin{equation}
\vcenter{\hbox{\includegraphics[scale=0.5]{pimiddle2.pdf} }}=-2Z_4\lambda_4\,\mathcal{A}_{k,2}\,,
\end{equation}
which prove the formula \eqref{prop1}.\\

\noindent
$\bullet$ \textit{6-point effective vertices.} From lemma \ref{propmelonfaces}, the external edges have to be hooked to three vertices of the same type, sharing the three external faces of the same colors running through the interior of the diagrams building $\bar{\pi}_k^{b_3^{(1)}}$. We then expect the following structure:
\begin{equation}
\bar{\pi}_k^{b_3^{(1)}}=\vcenter{\hbox{\includegraphics[scale=0.4]{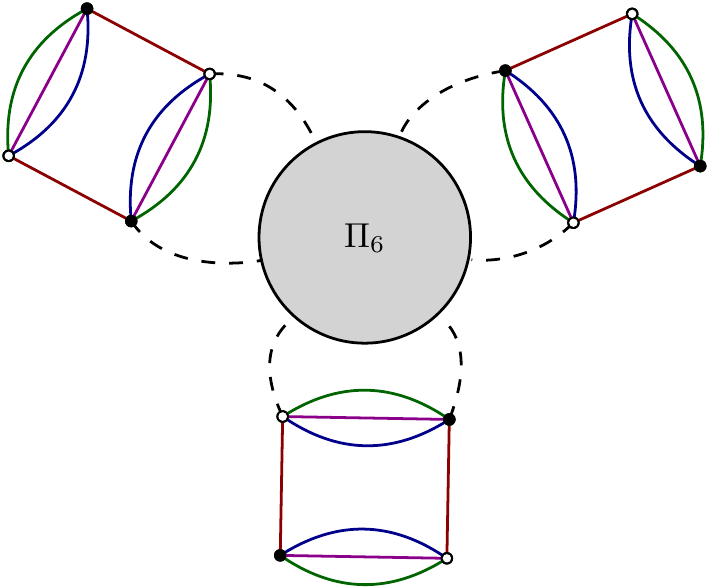}}}\,.
\end{equation}
As for $4$-points graphs, the kernel $\Pi_6$ may be decomposed into one particle irreducible components as:
\begin{equation}
\vcenter{\hbox{\includegraphics[scale=0.4]{structuresixpoints.pdf}}}=\vcenter{\hbox{\includegraphics[scale=0.4]{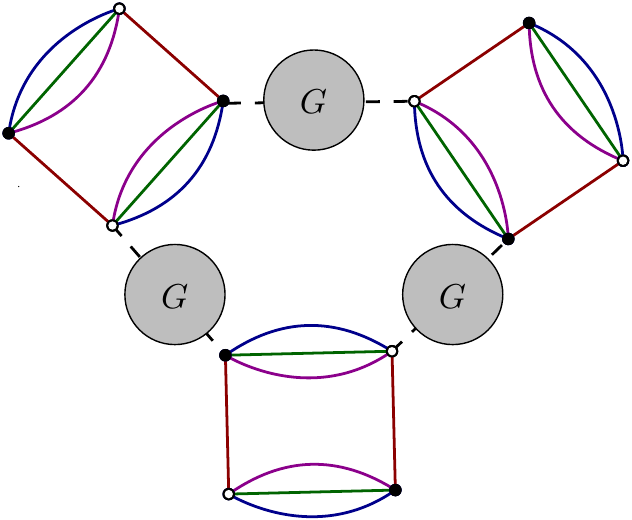}}}+\,\vcenter{\hbox{\includegraphics[scale=0.4]{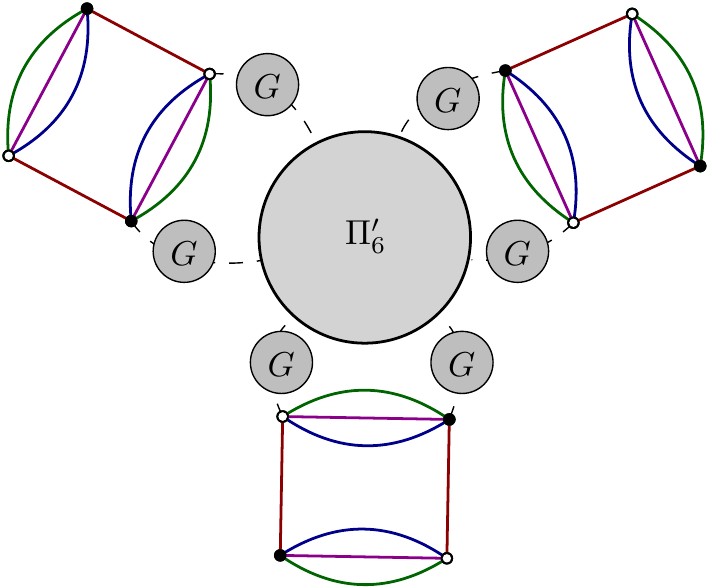}}}\,.
\end{equation}
The procedure may be conduct recursively like for the $4$-point functions. The perturbative expansion of the connected one particle irreducible component $\Pi_6^\prime$ start without vertex, and a moment of reflection show that it is nothing but $\Pi_6$ itself. Solving the recurrence, we get that each of the three arms reconstruct the structure of the effective $4$-point function, as a result:
\begin{equation}
\bar{\pi}_k^{b_3^{(1)}}=\vcenter{\hbox{\includegraphics[scale=0.4]{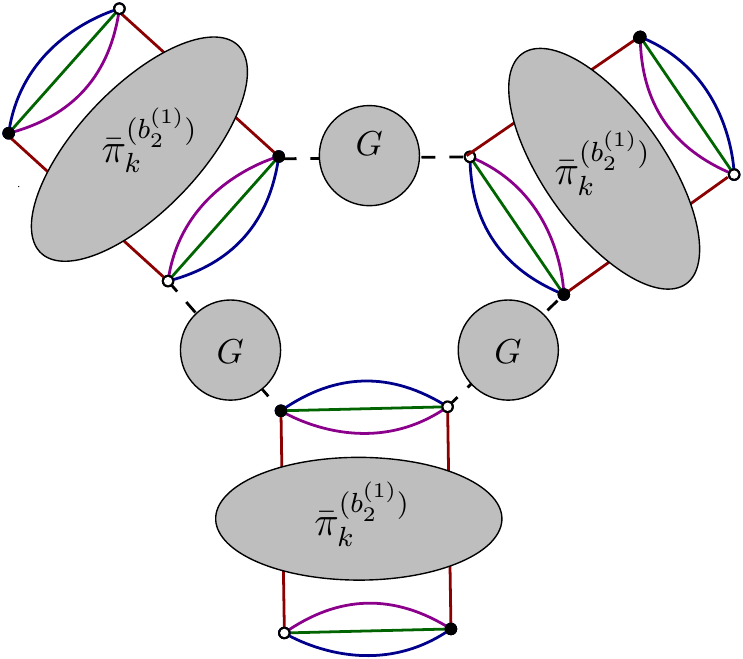}}}=K\,\big(\bar{\pi}_k^{(b_2^{(i)})}\big)^3 \,\mathcal{A}_{k,3}\,,\label{structure6}
\end{equation}
where the grey bubbles indexed with $\bar{\pi}_k^{b_2^{(1)}}$ represent effective $4$-point functions, and where for the last term, we translate the diagram into formula. The remaining numerical factor $K$ may be fixed from the leading order in the perturbative expansion. This term involves $3$ vertices, each of them having two orientations, then we have to $K=2^4/3!=8/3$. \\

\noindent
$\bullet$ \textit{8-point effective vertices.} As for $4$ and $6$ point functions, because of lemma \ref{propmelonfaces} we expect the following configuration for the boundary vertices:
\begin{equation}
\bar{\pi}_k^{b_4^{(1)}}=\vcenter{\hbox{\includegraphics[scale=0.4]{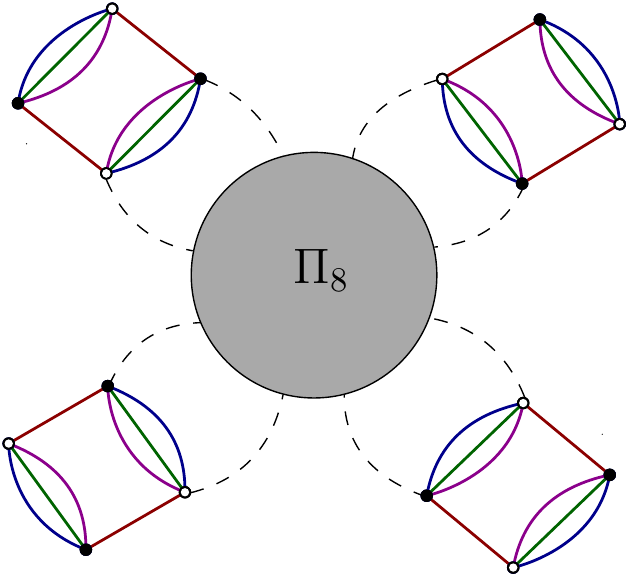}}}\,.
\end{equation}
Decomposing the kernel $\Pi_8$ following one particle irreducible parts, we get:
\begin{align}
\vcenter{\hbox{\includegraphics[scale=0.4]{structuresixpoints10.pdf}}}&=\vcenter{\hbox{\includegraphics[scale=0.45]{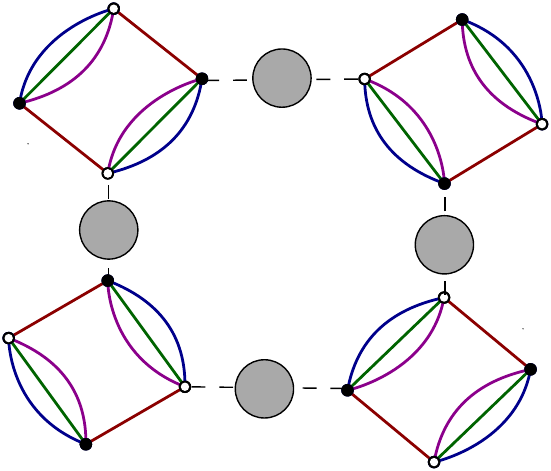}}}+\vcenter{\hbox{\includegraphics[scale=0.4]{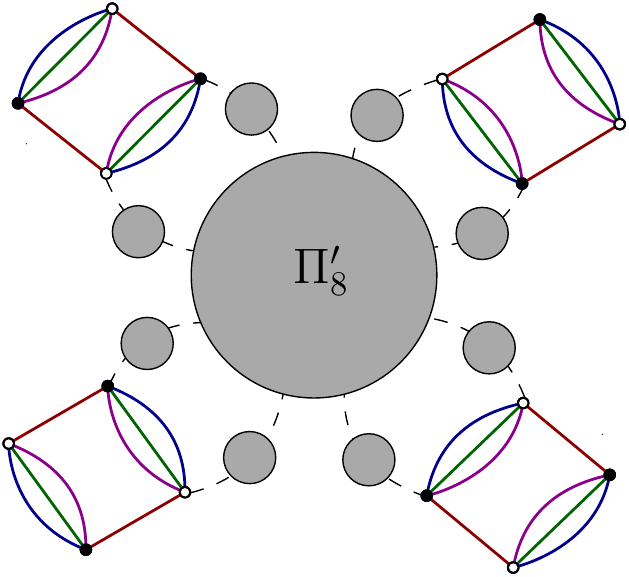}}}+\vcenter{\hbox{\includegraphics[scale=0.4]{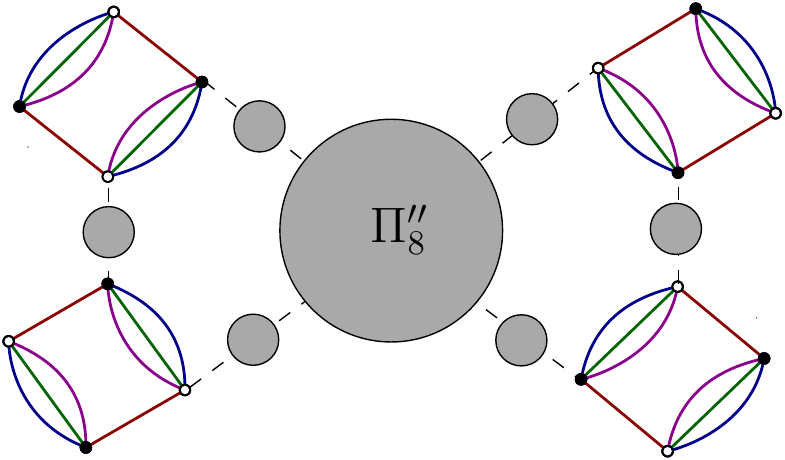}}}\,.
\end{align}
Note that $\Pi_8^{\prime\prime}$ is a one-particle irreducible function, at least of order one; and it is not hard to check that it is nothing but $\pi_k^{(b_2^{(1)})}$. Recursively, we get, as for $6$-point function:
\begin{equation}
\bar{\pi}_k^{b_4^{(1)}}=\vcenter{\hbox{\includegraphics[scale=0.45]{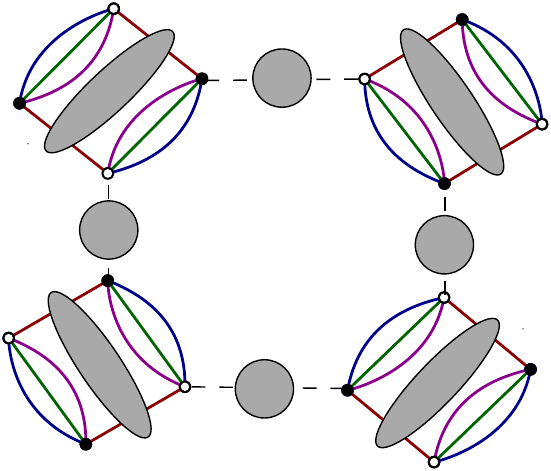}}}+\vcenter{\hbox{\includegraphics[scale=0.4]{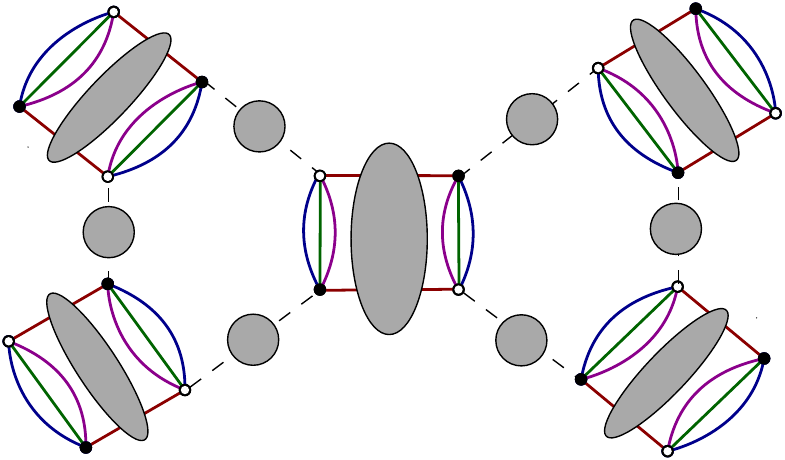}}}=-\frac{2^33!}{4!}\big(\bar{\pi}_k^{(b_2^{(i)})}\big)^4 \,\mathcal{A}_{k,4}+\frac{2^7}{4!}\big(\bar{\pi}_k^{(b_2^{(i)})}\big)^5\,\big(\mathcal{A}_{k,3}\big)^2\,.
\end{equation}

\begin{flushright}
$\square$
\end{flushright}

\noindent
The structure equations in the full sector, including $6$-point vertices must be easily deduced from proposition \ref{prop4}. Let us recall that $6$-point vertices may be obtained as connected sum of two $4$-point vertices using the following definition:
\begin{definition}
Let $b_2^{(i)}$ and $b_2^{(j)}$ two $4$-point bubbles. Let $n\in b_2^{(i)}$ and $\bar{n}\in b_2^{(j)}$ two white and black nodes and $e$ a dotted edge joining together there two nodes. The connected sum $b_2^{(i)}\sharp_{n\bar{n}}b_2^{(j)}$ is defined as the bubble obtained from the two following successive moves:
\begin{itemize}
\item Deleting the edge $e$
\item Connecting together the colored edges hooked to $n$ and $\bar{n}$ following their respective colors
\end{itemize}
\end{definition}
Figure \ref{figcont} below provides an illustration.
\begin{center}
\begin{equation*}
\vcenter{\hbox{\includegraphics[scale=0.6]{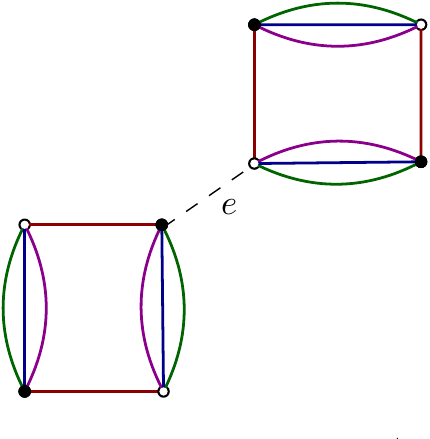} }}\,\underset{/e}{\longrightarrow} \,\vcenter{\hbox{\includegraphics[scale=0.8]{Melon2.pdf} }}
\end{equation*}
\captionof{figure}{Connected sum of two quartic bubbles.}\label{figcont}
\end{center}
The leading order graphs in the full sector, including $6$-point vertices can then be obtained from the leading order graphs in the restricted quartic sector, contracting some edges between adjacent quartic vertices. From definition, the contracted edges do not change the number of faces; moreover, each contraction do not change the power counting : deleting an edge increase the power counting of $2$, compensating exactly the lost of two quartic vertices, each of them with canonical dimension $1$. \\

\noindent
Let us consider a 1PI $4$-point graphs contributing to $\Gamma^{(4)}_k$, and $e$ an edge between two adjacent vertices. We have to distinguish two cases:
\begin{itemize}
\item The edge $e$ is including into an effective propagator $G_k$.
\item The edge $e$ is between two consecutive vertices on the chain.
\end{itemize}
In the first case, the resulting $2$-point subgraph is nothing but a contribution to the full $2$ point function. In the second case, the contraction of the edge generate a $6$-point graph along the chain as follows:
\begin{equation}
\vcenter{\hbox{\includegraphics[scale=0.8]{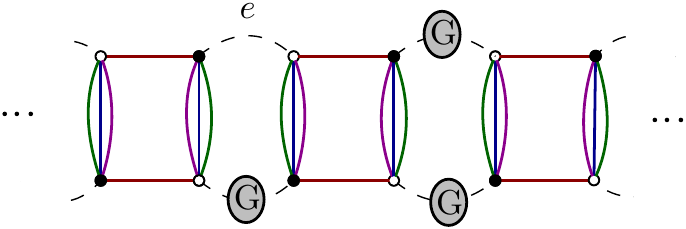} }} \,\underset{/e}{\longrightarrow} \, \vcenter{\hbox{\includegraphics[scale=0.8]{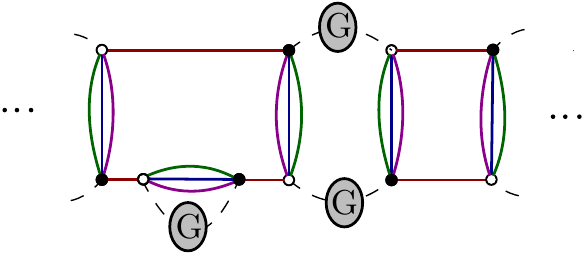} }}\,.
\end{equation}
Two elementary quartic vertices have been replaced by an effective quartic vertex, building from an elementary $6$-point vertex by contracting two external fields with the effective propagator $G_k$ to form an effective $3$-dipole. Note that the propagator $G$ is becoming the full effective propagator, including $6$ point vertices as well. We call \textit{type--2 quartic vertices} these new building blocks in the chain;
\begin{equation}
\vcenter{\hbox{\includegraphics[scale=0.8]{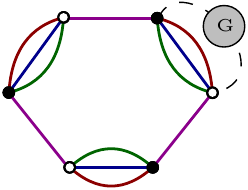} }}=\vcenter{\hbox{\includegraphics[scale=0.8]{Melon1.pdf} }}\times\vcenter{\hbox{\includegraphics[scale=0.8]{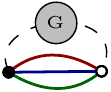} }}
\end{equation}
and their explicit expression may be easily computed. For zero-external momenta, we get:
\begin{equation}
\vcenter{\hbox{\includegraphics[scale=0.8]{Melontype21.pdf} }}=Z_6\lambda_6\times \mathfrak{b}_k\,,\qquad\mathfrak{b}_k:=3\sum_{\vec{p}\in\mathbb{Z}^{\droit-1}}\,G_k(\vec{p}\,)
\end{equation}
where $Z_6\lambda_6$ denote the bare $6$-point coupling, and the factor $3$ in the definition of $\mathfrak{b}_k$ comes from the counting of all possibles contractions. The structure of the full chain, including $6$-point vertices may be deduced exactly following the same strategy as for the proof of proposition \ref{prop4}. Denoting as $\tilde{\pi}^{(b_2^{(1)})}_{k}$ the leading order $4$-point effective kernel made of chains building only of type--2 quartic vertices, we get:
\begin{equation}
\tilde{\pi}^{(b_2^{(1)})}_{k}=\frac{Z_6\lambda_6\,\mathfrak{b}_k}{1+2Z_6\lambda_6 \mathfrak{b}_k \mathcal{A}_{k,2}}\,.
\end{equation}
To find the full effective vertex, we have to take into account the proliferation of the two type of $4$-point vertices along the chain. There are three cases that we have to distinguish:
\begin{itemize}
\item The boundary vertices are both of type--1
\item The boundary vertices are both of type--2
\item The boundary vertices are assorted
\end{itemize}
For the first case, starting with $\bar{\pi}^{(b_2^{(1)})}_{k}$, we may consider all the possible insertion of $\tilde{\pi}^{(b_2^{(1)})}_{k}$ without breaking the boundary constraint. We get:
\begin{equation}
\bar{\pi}^{(b_2^{(1)})}_{k}+\bar{\pi}^{(b_2^{(1)})}_{k}\tilde{\pi}^{(b_2^{(1)})}_{k}\bar{\pi}^{(b_2^{(1)})}_{k}+\bar{\pi}^{(b_2^{(1)})}_{k}\tilde{\pi}^{(b_2^{(1)})}_{k}\bar{\pi}^{(b_2^{(1)})}_{k}\tilde{\pi}^{(b_2^{(1)})}_{k}\bar{\pi}^{(b_2^{(1)})}_{k}+\cdots=\frac{\bar{\pi}^{(b_2^{(1)})}_{k}}{1-\tilde{\pi}^{(b_2^{(1)})}_{k}\bar{\pi}^{(b_2^{(1)})}_{k}}\,.
\end{equation}
Proceeding in the same way for the remaining cases, we get for the full effective vertex:
\begin{equation}
{\pi}^{(b_2^{(1)})}_{k}=\frac{\bar{\pi}^{(b_2^{(1)})}_{k}+\tilde{\pi}^{(b_2^{(1)})}_{k}+\bar{\pi}^{(b_2^{(1)})}_{k}\tilde{\pi}^{(b_2^{(1)})}_{k}}{1-\tilde{\pi}^{(b_2^{(1)})}_{k}\bar{\pi}^{(b_2^{(1)})}_{k}}\,.
\end{equation}
The structure of the $6$-point effective vertex may be deduced in the same way. From equation \eqref{structure6}, we know that the structure equation for $6$-point effective vertices in the quartic sector is an effective loop build of three effective propagators, hooked together effective $4$-point vertices. To find the contributions involving bare $6$-point couplings, we have to list all the possible contractions of an edge $e$ into a typical graph contributing to $\bar{\pi}_k^{b_3^{(1)}}$, in order to build a connected sum of two $4$-point bare couplings. We have three different ways :
\begin{itemize}
\item the edge $e$ is on the effective loop, between two adjacent effective $4$-point vertices
\item the edge $e$ is on one of the three effective propagators building the central effective loop
\item the edge $e$ is on one of the effective $4$-point vertices
\end{itemize}
Once again, for the last cases, the resulting $2$-point subgraph provides a contribution to the full $2$-point function. In the same way, the resulting $4$-point subgraph provides a contribution to the full $4$-point function. None of these two moves modify the global structure of the graph: It remains built of three effective $4$-point vertices connected together with $2$-point graphs to form an effective loop of length three. The first move however modifies the global structure. It generates a $6$-point vertex to which three hands building effective $4$-point graphs are hooked. Listing all the allowed configurations, distinguished from their respective symmetry factors, we get:
\begin{equation}
{\pi}_k^{b_3^{(1)}}=\frac{8}{3}\big({\pi}_k^{(b_2^{(1)})}\big)^3 \,\mathcal{A}_{k,3}+\tilde{\pi}_k^{b_3^{(1)}}\,,
\end{equation}
where at this stage $\mathcal{A}_{k,3}$ is not restricted on the quartic sector, and where, graphically :
\begin{align}
\tilde{\pi}_k^{b_3^{(1)}}=\vcenter{\hbox{\includegraphics[scale=0.4]{Melon2.pdf} }}&+\, \vcenter{\hbox{\includegraphics[scale=0.4]{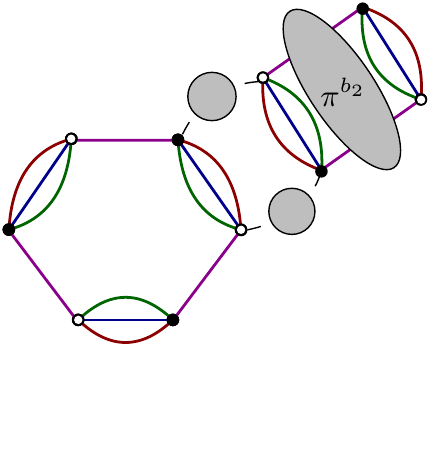} }}+\vcenter{\hbox{\includegraphics[scale=0.4]{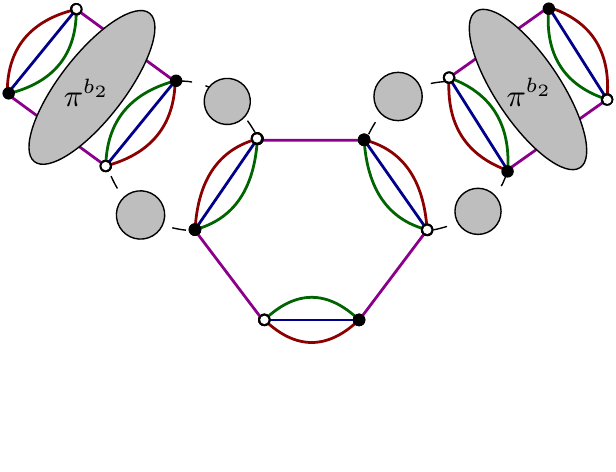} }}+\vcenter{\hbox{\includegraphics[scale=0.4]{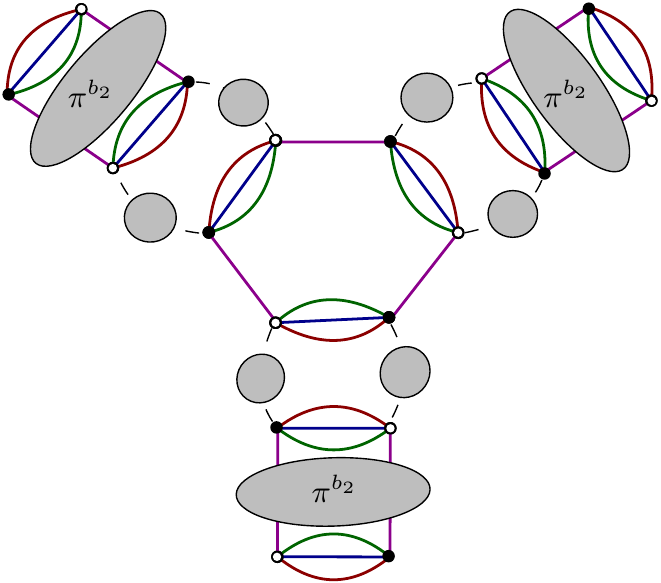} }}\,.\label{defphi61}
\end{align}
The relative symmetry factors for each contribution can be easily computed considering the leading order contributions in the perturbative expansion, and we get:
\begin{equation}
\tilde{\pi}_k^{b_3^{(1)}}=Z_6\lambda_6\left(1-6{\pi}_k^{(b_2^{(i)})}\mathcal{A}_{k,2}+12\big({\pi}_k^{(b_2^{(i)})}\mathcal{A}_{k,2}\big)^2-16\big({\pi}_k^{(b_2^{(i)})}\mathcal{A}_{k,2}\big)^3\right)\,.\label{barphi6}
\end{equation}
As a result, from the renormalization conditions ${\pi}_k^{b_3^{(1)}}=\lambda_6(k)$ and ${\pi}_k^{b_2^{(1)}}=\lambda_4(k)$, we then deduce the relation between the bare $6$-point coupling and the effective renormalizable couplings:
\begin{equation}
Z_6\lambda_6=\dfrac{\lambda_6(k)-\frac{4}{3}\lambda_4^3(k) \,\mathcal{A}_{k,3}}{1-6\lambda_4(k)\mathcal{A}_{k,2}+12\big(\lambda_4(k)\mathcal{A}_{k,2}\big)^2-16\big(\lambda_4(k)\mathcal{A}_{k,2}\big)^3}\,. \label{8decomp}
\end{equation}
Finally, it is easy to check that the $8$-point kernel $\pi^{b_4^{(1)}}$ decomposes as:
\begin{equation}
\pi^{b_4^{(1)}}=-2\big(\bar{\pi}_k^{(b_2^{(i)})}\big)^4\mathcal{A}_{k,4} +\tilde{\pi}^{b_4^{(1)}}\,,
\end{equation}
where graphically, the last component $\tilde{\pi}^{b_4^{(1)}}$ is given by:
\begin{equation}
\tilde{\pi}^{b_4^{(1)}}=\vcenter{\hbox{\includegraphics[scale=0.4]{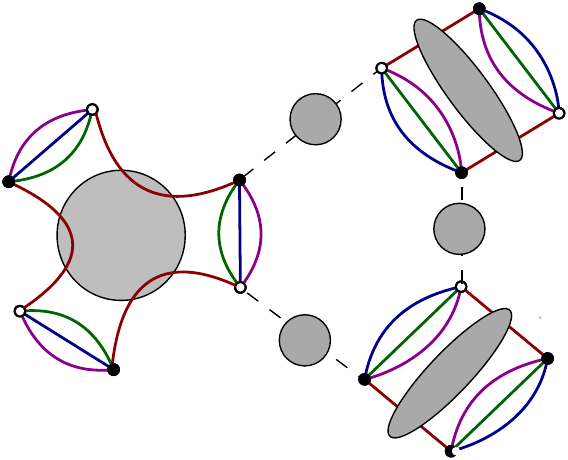} }}+\vcenter{\hbox{\includegraphics[scale=0.4]{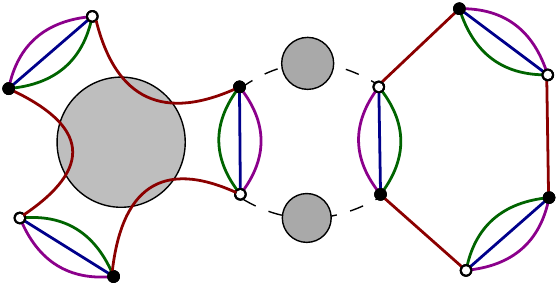} }}+\vcenter{\hbox{\includegraphics[scale=0.4]{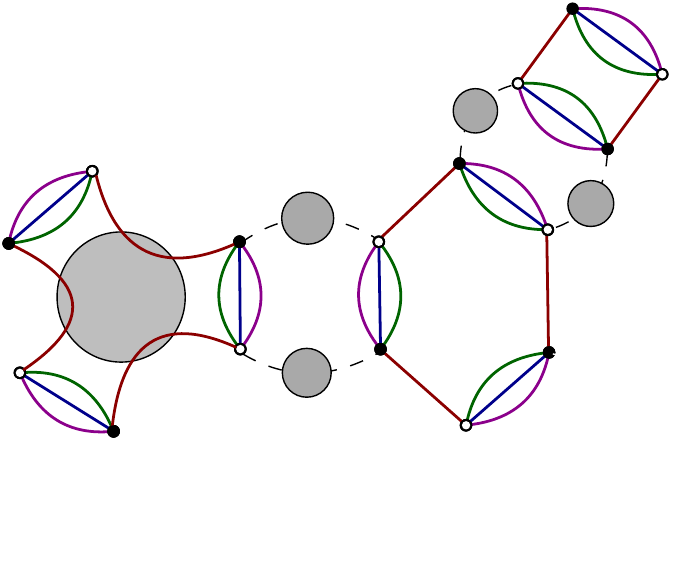} }}+\vcenter{\hbox{\includegraphics[scale=0.4]{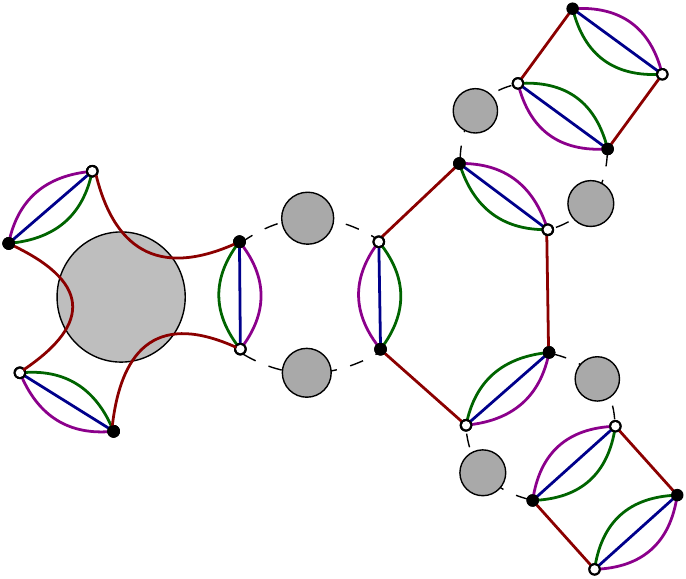} }}\,\label{8diags}
\end{equation}
where the effective $6$-point vertex has been pictured as a grey disk labeled with $\pi^{(b_3)}$, the boundary structure being explicit. Note we added into $\tilde{\pi}^{b_4^{(1)}}$ the second contribution in \eqref{8decomp}, to build the first contribution in \eqref{8diags}. Computing the relative symmetry factors, using formula \eqref{barphi6} to express the bare $6$-point couplings in term of the effective couplings at scale $k$, we get finally :
\begin{align}
\nonumber\tilde{\pi}^{b_4^{(1)}}= \lambda_6(k) \bigg(24\mathcal{A}_{k,3}\lambda_4^2(k) -&\frac{9}{2}\dfrac{\lambda_6(k)-\frac{4}{3}\lambda_4^3(k) \,\mathcal{A}_{k,3}}{1-6\lambda_4(k)\mathcal{A}_{k,2}+12\big(\lambda_4(k)\mathcal{A}_{k,2}\big)^2-16\big(\lambda_4(k)\mathcal{A}_{k,2}\big)^3}\\
&\qquad\qquad\times \mathcal{A}_{k,2}\left(1-4\lambda_4(k)\mathcal{A}_{k,2}+4(\lambda_4(k)\mathcal{A}_{k,2})^2\right)\bigg)\,,\label{phi8}
\end{align}
which close the system \eqref{system1}. The remaining piece required to complete the set of equation is an equation for $Z$, or equivalently for the anomalous dimension, that we will fix on the next section.

\subsection{Anomalous dimension and Ward identities} \label{section anomalous}
Following standard definition in the symmetric phase, the \textit{wave function normalization} $Z(k)$ and the \textit{anomalous dimension} $\eta(k)$, completing the set of power-counting renormalizable operators, are defined as:
\begin{definition} \label{defanomalous}\textbf{Wave function renormalization and anomalous dimension.} In the symmetric phase, the wave function renormalization is the flow-dependents weight of the laplacian term is the derivative expansion of the $2$-point function $\Gamma^{(2)}_k$:
\begin{equation}
Z(k):=\dfrac{d}{dp_1^2}\Gamma^{(2)}_k(\vec{p}=\vec{0}\,)\,.
\end{equation}
The anomalous dimension $\eta(k)$ is then defined as:
\begin{equation}
\eta(k):=\frac{d}{dk}\ln(Z(k))\,.
\end{equation}
\end{definition}
Note that outside of the symmetric phase, i.e. if we expand around a non-zero vacuum, an additional dependence of $Z$ on the mean fields may be expected. A standard approximation, already discussed in \cite{Lahoche:2018oeo} is to keep only the leading order terms of the derivative expansion (DE) for the computation of the loop integrals $L_j(p)$ involved in the system \eqref{system1}. More precisely:
\begin{definition} \label{LDE}\textbf{Leading DE approximation.} In the relevant windows of momenta allowed by the distribution $\dot{r}_k(\vec{p}\,)$, the renormalized effective $2$-point function $\Gamma^{(2)}_k(\vec{p}\,)$ is approximated by the two first terms of the derivative expansion:
\begin{equation}
\Gamma^{(2)}_k(\vec{p}\,)=m^2(k)+Z(k)\vec{p}\,^2\,.\label{LDEequ}
\end{equation}
\end{definition}
In order to get fixed points, we will consider the \textit{dimensionless and renormalized} couplings, extracting the flow dependence coming from proper canonical dimension and wave function renormalization:
\begin{definition}\label{dimensionless} \textbf{Renormalized and dimensionless couplings.} In the deep UV ($k\gg 1$) the renormalized and dimensionless couplings are defined as:
\begin{equation}
m^2=Z(k)k^2\bar{m}^2\,,\qquad \lambda_4=Z^2(k)k\bar{\lambda}_4\,,\qquad \lambda_6=Z^3(k)\bar{\lambda}_6\,.
\end{equation}
\end{definition}
This is the RG flow equations for these couplings that we have to compute. To this end, we have to get the flow equation for $Z(k)$. From definition \ref{defanomalous}, we obtain:
\begin{equation}
\dot{Z}(k)=-2\bigg( \frac{d\pi_k^{(b_2^{(i)})}}{dp^2}(p=0) \bigg)L_1(0)-2\lambda_4(k)\bigg( \frac{dL_1}{dp^2}(p=0) \bigg) \,.\label{dotZ}
\end{equation}
The second piece of the right hand side may be computed from the definition \ref{LDE}, once chosen the regulator function $r_k(\vec{p}\,)$. In order to get a tractable closed system, the first term of the right hand side have to be computed in term of the other couplings as well. WT identities comes from the Unitary invariance of the tensorial interactions and the formal translation invariance of the Lebesgue measure in the path integral \eqref{Z}. Considering an infinitesimal unitary transformation along a single axis on the torus $\U(1)^d$, and following the proof given in \cite{Lahoche:2018vun}, we may deduce the following theorem:
\begin{theorem}
The partition function $\mathcal{Z}_k[J,\bar{J}]=:e^{W_k[J,\bar{J}]}$ of the theory defined by the equation \eqref{Z} verifies the WT identity:
\bea\label{left}
\sum_{\vec{p}_\bot, \vec{p}_\bot\,^{\prime}} \prod_{j\neq 1} \delta_{p_jp_j^\prime} \bigg\{\big[C_s^{-1}(\vec{p}\,^{2})-C_s^{-1}(\vec{p}\,^{\prime\,{2}})\big]\left[\frac{\partial^2 W_s}{\partial \bar{J}_{\vec{p}\,^\prime}\,\partial {J}_{\vec{p}}}+\bar{M}_{\vec{p}}M_{\vec{p}\,^\prime}\right]-\bar{J} _{\vec{p}}\,M_{\vec{p}\,^\prime}+{J} _{\vec{p}\,^\prime}\bar{M}_{\vec{p}}\bigg\}=0\,,\label{Ward0}
\eea
with $\vec{p}_\bot:=(0,p_2,\cdots ,p_\droit)\in \mathbb{Z}^{\droit}$ and the means fields $M$ and $\bar{M}$ defined from equations \eqref{M}.
\end{theorem}
Deriving twice with respect to $M$ and $\bar{M}$, and vanishing all the odd vertex functions, a required in the symmetric phase, we then deduce a relation between $6$-point and $4$-points effective functions. In particular, the dependence with respect to the $4$-point functions involves a difference between functions having different momenta. Taking the continuum limit, and setting to zero the external momenta, we then deduce the following relation (see Appendix A of \cite{Lahoche:2018vun}):
\begin{corollary} \textbf{Second zero momenta WT--identity:}\label{secondWI}
In the symmetric phase, the zero-momenta $6$ and $4$--point functions satisfy:
\bea
Z_{-\infty}\Big[\frac{3}{2}\pi^{(b_3)}\mathcal{L}_1-4(\pi^{(b_2)})^2\mathcal{L}_2\Big]=-\frac{d}{dp^2}\pi^{(b_2)}(p=0)\,.\label{sixpointWI}
\eea
where $Z_{-\infty}$ denotes the wave-function counter-term, canceling the UV divergences and computed for $k=0$; and where we defined the loop functions $\mathcal{L}_j$ as:
\begin{equation}
\mathcal{L}_j:=\sum_{\vec{p}_\bot}\bigg(1+\frac{\partial \tilde{r}_{s}(\vec{p}_{\bot})}{\partial p_1^2}\bigg)[G_s(\vec{p}_\bot)]^{j+1}\,,\qquad r_s=:Z_{-\infty}\tilde{r}_s\,.
\end{equation}
\end{corollary}
This relation link the derivative of the $4$-point effective vertex with the $6$ and $4$-point vertices. However to be
exploitable, this relation have to be completed with the WT identity obtained from \eqref{Ward0}, deriving one time with respect to $M$ and $\bar{M}$ and setting to zero the external momenta:
\begin{corollary}\textbf{First zero momenta WT--identity}\label{FirstWI}
In the symmetric phase, the zero-momenta $4$-point fonction satisfies:
\begin{equation}
2 \pi^{(b_2)}Z_{-\infty}\mathcal{L}_1=-(Z-Z_{-\infty})\,.\label{equationWTI}
\end{equation}
\end{corollary}
To obtain this relation, we used of the definition \eqref{defanomalous}. Note that in the equations \eqref{sixpointWI} and \eqref{equationWTI}, the derivative of the bare covariance $C_s^{-1}(\vec{p}\,^{2})$ does not involves ‘‘boundary terms" coming from UV cut-off. This is because, to derive these equations, we chose to regularize the divergent integrals using \textit{dimensional regularization}, exploiting the analytic properties of the integrals with respect to the group dimension $D$. The \textit{continuum limit} in this case is then defined when $D\to1$. \\

\noindent
For our purpose, the interest of the Ward identity \eqref{equationWTI} is that it allows to express $Z_{-\infty}$ in terms of effective quantities at scale $k$:
\begin{equation}
Z_{-\infty}=\frac{Z(k)}{1-2\mathcal{L}_1\lambda_4(k)}\,. \label{approx1}
\end{equation}
Let $\mathcal{L}_j=:\mathcal{A}_{k,j+1}+\frac{1}{Z_{-\infty}}\Delta_j(k)$. Because of the derivative of the regulator $\partial r_k/\partial p^2$ into the last term $\Delta_j(k)$, one expect that it can be computed using the LDE approximation given by definition \eqref{LDE}, as well as the functions $L_j(p)$ defined in \eqref{sum} and involved into the flow equations \eqref{system1}. In particular, for the choice of the \textit{modified Litim regulator} \cite{Litim:2000ci}-\cite{Litim:2001dt}:
\begin{equation}
r_k(\vec{p}\,)=Z(k)(k^2-\vec{p}\,^2)\theta(k^2-\vec{p}\,^2)\,,\label{litim}
\end{equation}
the allowed windows of momenta is the same for $\dot{r}_k$ and $\partial r_k/\partial p^2$ -- $\theta(x)$ denoting the standard Heaviside step function. In the rest of this paper, we restrict our attention on this choice, extensively used in the tensor literature, and showed to be optimal \cite{Litim:2000ci}-\cite{Litim:2001dt} in some practical applications. In the continuum limit, replacing the sums by the corresponding well defined integrations, we get for $\Delta_j$ and $L_j$:
\begin{align}
\Delta_j(k)&= -\frac{Z(k)\Omega_3}{(Zk^2+m^2)^{j+1}}k^{3} \\
L_j(p)&=Z(k)\Omega_3(k^2-p^2)^{\frac{3}{2}}\,\frac{2k^2+\frac{2}{5}\eta(k)(k^2-p^2) }{(k^2Z(k)+m^2(k))^{j+1}}\label{approx2}
\end{align}
where $\Omega_3=\frac{4\pi}{3}\approx 4.19$ denotes the volume of the $3$-ball. The computation of the last piece $\mathcal{A}_{k,2}$ is a little bit subtle, the momentum integration being over the full momentum space:
\begin{equation}
\mathcal{A}_{k,j+1}=\sum_{\vec{p}_\bot\in\mathbb{Z}^3} \frac{\theta(k^2-\vec{p}\,^2_\bot)}{(\Gamma^{(2)}(\vec{p}_\bot)+r_k(\vec{p}_\bot))^{j+1}}+\sum_{\vec{p}_\bot\in\mathbb{Z}^3} \frac{\theta(\vec{p}\,^2_\bot-k^2)}{(\Gamma^{(2)}(\vec{p}_\bot))^{j+1}}\,.
\end{equation}
Once again, the first term on the right hand side can be computed using the LDE approximation, the allowed windows of momenta being the same as for $r_k$. For the second term however, the use of this approximation would be an additional approximation, beyond of the domain where it can be justified. In \cite{Lahoche:2018oeo}, we have showed explicitly that this approximation applied for superficially divergent integral leads to bad defined results, especially because it introduce a dependence on $Z(k)$ in the deep UV, far away from the scale $k$. However, in the same discussion, the authors show that for the superficially convergent integrals, this approximation may be not so bad, because none additional dependence on $k$ is expected on the deep UV. In other words, we assume that the smooth cut-off arising from the convergence of the integrals allows to use the LDE approximation. We get explicitly:
\begin{equation}
\mathcal{A}_{k,j+1}= \frac{\Omega_3k^3}{(Z(k)k^2+m^2(k))^{j+1}}+3\Omega_3 \int_k^\infty \frac{p^2 dp}{(Zp^2+m^2)^{j+1}}\,.
\end{equation}
For any quantity $X(\{\lambda_i\},m^2)$ depending on the couplings, we denote by $\bar{X}(\{\bar{\lambda}_i\},\bar{m}^2)$ the corresponding renormalized and dimensionless quantity:
\begin{equation}
X(\{\lambda_i\},m^2)=k^{d_X} Z^{q_x} \bar{X}(\{\bar{\lambda}_i\},\bar{m}^2)\,.
\end{equation}
Defining
\begin{equation}
\mathcal{I}_j(\bar{m}^2):=\int_1^\infty \frac{x^2 dx}{(x^2+\bar{m}^2)^{j+1}}\,,
\end{equation}
we get:
\begin{equation}
\bar{\mathcal{A}}_{k,j+1}= -\bar{\Delta}_j+3\Omega_3\mathcal{I}_j\,,\quad \bar{\Delta}_j= -\frac{\Omega_3}{(1+\bar{m}^2)^{j+1}}\,,
\end{equation}
and the equation \eqref{approx1} writes as:
\begin{equation}
Z_{-\infty}=Z(k)\frac{1-2\bar{\lambda}_4(k)\frac{\Omega_3}{(1+\bar{m}^2)^{2}}}{1-2\bar{\lambda}_4(k)\bar{\mathcal{A}}_{k,2}}\,. \label{approx3}
\end{equation}
Introducing this expression into the relation \eqref{sixpointWI}, we obtain:
\begin{equation}
\overline{\frac{d\pi^{(b_2)}}{dp^2}}(p=0)=-3\Omega_3\frac{\frac{3}{2}\bar{\lambda}_6\,\mathcal{I}_1(\bar{m}^2)-4\bar{\lambda}_4^2\left(\mathcal{I}_2(\bar{m}^2)-\frac{2\bar{\lambda}_4\Omega_3}{(1+\bar{m}^2)^{2}}\mathcal{I}_2(\bar{m}^2)\right)-\frac{\bar{\lambda}_4\bar{\lambda}_6\Omega_3}{(1+\bar{m}^2)^2}\mathcal{I}_1(\bar{m}^2)}{1-2\bar{\lambda}_4\bar{\mathcal{A}}_{k,2}} \,.\label{equationderiv}
\end{equation}
From equation \eqref{equationderiv} and \eqref{approx2}, and from definition \eqref{defanomalous} giving the anomalous dimension $\eta(k)$, the equation \eqref{dotZ} for $\dot{Z}$ becomes:
\begin{equation}
\eta=-2\overline{\frac{d\pi^{(b_2)}}{dp^2}}\Omega_3\,\frac{2+\frac{2}{5}\eta }{(1+\bar{m}^2)^{2}} +2\bar{\lambda}_4\Omega_3 \,\frac{3+\eta }{(1+\bar{m}^2)^{2}}\,.\label{equationeta}
\end{equation}
$\eta $ can be extracted simply as
\begin{equation}
\boxed{
\eta=2\Omega_3\frac{3\bar{\lambda}_4-2\overline{\frac{d\pi^{(b_2)}}{dp^2}}}{(1+\bar{m}^2)^{2}-2\bar{\lambda}_4\Omega_3+\frac{4}{5}\overline{\frac{d\pi^{(b_2)}}{dp^2}} \Omega_3 }\,.\label{anomalousflow}
}
\end{equation}
Using the definition \ref{dimensionless}, we deduce the flow equations for dimensionless and renormalized couplings. We summarize all these results in the following proposition:
\begin{proposition}
In the deep UV limit $k\gg1$, denoting as $\beta_2:=\dot{\bar{m}}^2$, $\beta_4:=\dot{\bar{\lambda}}_4$ and $\beta_6:=\dot{\bar{\lambda}}_6$, we have:
\begin{align}
\beta_2&=-(2+\eta)\bar{m}^2-2\Omega_3\bar{\lambda}_4 \frac{2+\frac{2}{5}\eta }{(1+\bar{m}^2)^{2}}\,,\label{system21}\\
\beta_4&=-(1+2\eta)\bar{\lambda}_4-3\bar{\lambda}_6\Omega_3 \frac{2+\frac{2}{5}\eta }{(1+\bar{m}^2)^{2}}+4\Omega_3\bar{\lambda}_4^2 \,\frac{2+\frac{2}{5}\eta }{(1+\bar{m}^2)^{3}}\,,\label{system22}\\
\beta_6&=-3\eta\bar{\lambda}_6-4\Omega_3 \overline{\pi_k^{(b_4^{(1)})}} \frac{2+\frac{2}{5}\eta }{(1+\bar{m}^2)^{2}}+12\Omega_3 \bar{\lambda}_4\bar{\lambda}_6 \frac{2+\frac{2}{5}\eta }{(1+\bar{m}^2)^{3}}-8\Omega_3\bar{\lambda}_4^3 \frac{2+\frac{2}{5}\eta }{(1+\bar{m}^2)^{4}}\,,\label{system23}
\end{align}
where $\overline{\pi_k^{(b_4^{(1)})}}$ is given from the structure equation \eqref{phi8} as:
\begin{align}
\overline{\pi_k^{(b_4^{(1)})}}=-2\bar{\lambda}_4^4\bar{\mathcal{A}}_{k,4}+ 3\bar{\lambda}_6 \bigg(4\bar{\lambda}_4^2\bar{\mathcal{A}}_{k,3} -&\frac{3\bar{\mathcal{A}}_{k,2}}{2}\dfrac{\left(\bar{\lambda}_6-\frac{4}{3}\bar{\lambda}_4^3 \,\bar{\mathcal{A}}_{k,3}\right)\left(1-4\bar{\lambda}_4\bar{\mathcal{A}}_{k,2}+4(\bar{\lambda}_4\bar{\mathcal{A}}_{k,2})^2\right)}{1-6\bar{\lambda}_4\bar{\mathcal{A}}_{k,2}+12\big(\bar{\lambda}_4\bar{\mathcal{A}}_{k,2}\big)^2-16\big(\bar{\lambda}_4\bar{\mathcal{A}}_{k,2}\big)^3} \bigg)\,.
\end{align}
\end{proposition}

\section{Melonic phase space investigations}\label{sec4}
In this section, we investigate the phase space of the RG flow described with the system \eqref{system21}-\eqref{system23} in the vicinity of the Gaussian fixed point, where the constraint \eqref{constraint1} is trivially satisfied. however, beyond the Gaussian fixed point, we will show that this constraint is violated. Finally, we described the flow on the constrained subspace $\mathcal{E}$.

\subsection{Vicinity of the Gaussian fixed point}
Expanding the flow equations \eqref{system21}-\eqref{system23} in power of couplings, and keeping only the leading order contributions in power of couplings, we get a simplified system describing the flow in the vicinity of the Gaussian fixed point, where coupling vanish. For instance, in the exact formula \eqref{anomalousflow}, we retain ($\bar{\mathcal{I}}_1(0) =1$):
\begin{equation}
\eta\approx 6\Omega_3 \bar{\lambda}_4 +18\Omega_3^2\bar{\lambda}_6\,.\label{approxeta1}
\end{equation}
For our analysis however, we restrict our analysis to the one-loop contributions, which simplifies the incoming analysis. As a consequence, we discard the last term in the right hand side of the previous equation :
\begin{equation}
\eta\approx 6\Omega_3 \bar{\lambda}_4\,.\label{approxeta2}
\end{equation}
{With this approximation the relations \eqref{system21}-\eqref{system23} reduces to:
\begin{align}\label{system3 }
\beta_2&\approx-2\bar{m}^2-4\Omega_3 \bar{\lambda}_4\,,\\
\beta_4&\approx-\bar{\lambda}_4-6\Omega_3\bar{\lambda}_6\,,\\
\beta_6&\approx+\,6\Omega_3\bar{\lambda}_4\bar{\lambda}_6\,.
\end{align}
The same kind of equations has been obtained for all $\phi^6$-models studied in the literature, and extended discussions may be found in \cite{Carrozza:2014rba}-\cite{Carrozza:2017vkz},\cite{Lahoche:2016xiq},\cite{Carrozza:2014rya}. A qualitative phase portrait may be easily deduced from the system \eqref{system3 }. We split the vicinity of the Gaussian fixed point into four regions, $I$, $II$, $III$ and $IV$ as pictured on Figure \ref{RGGFP} below. The boundary between the regions $I$ and $II$ on the upper side, and between the regions $III$ and $IV$ on the lower side, corresponds to the line defined by the equation $\beta_4=0$. The line defined by the equation $\bar{\lambda}_6=0$ is the boundary between regions $I$ and $IV$ on one hand, and between regions $II$ and $III$ on the other hand.\\

\noindent
In the positive neighbourhood of the region $II$, where the couplings are both positives, and the beta functions $\beta_4$ and $\beta_6$ have opposite signs, $\bar{\lambda}_4$ decrease whereas $\bar{\lambda}_6$ increase, and the RG trajectory is then repelled from the horizontal axis and goes to the vertical axis. The maximum for $\bar{\lambda}_6$ arise for $\bar{\lambda}_4=0$, that is when the RG trajectory reaches the vertical axis. Because $\bar{\lambda}_6>0$, $\beta_4$ remains negative, so that $\bar{\lambda}_4$ pass through the vertical axis, and becomes negative. From now, $\beta_4$ and $\beta_6$ have the same signs, and the couplings decrease together. The coupling $\bar{\lambda}_6$ goes to zero, whereas the coupling $\bar{\lambda}_4$ becomes more negative. The trajectory then goes far away from the vertical axis, and closer to the horizontal axis. Let us focus on the region $\bar{\lambda}_6\geq 1$. In the vicinity of the vertical axis, $\bar{\lambda}_4$ remains very close to zero, and as soon as $\vert \bar{\lambda}_4\vert < 1$, the tangent vectors of the RG trajectory are almost horizontals. The more $\vert \bar{\lambda}_4\vert$ grows, the more the tangent vectors becomes vertical so that the trajectory go faster and faster to the horizontal axis until the trajectory touches the line $\beta_4=0$ and pass through, reaching the region $I$. In the region $I$, $\beta_6$ remains negative, but $\beta_4$ becomes positive, and then $\bar{\lambda}_4$ begins to grow. At this stage, stuck at the bottom with the red trajectory, the considered trajectory has to reach the Gaussian fixed point. \\

\noindent
On \ref{RGGFP}a we drawed a qualitative picture obtained from our previous analysis and \ref{RGGFP}b and \ref{RGGFP1} give the numerical analysis. In a sufficiently small neighborhood in the vicinity of the Gaussian fixed point, the sign of the coupling $\bar{\lambda}_6$ is then conserved along the flow, and the positivity of the initial $\phi^6$-action ensures that the starting point have to be in the upper plan, for positive $\bar{\lambda}_6$; and our discussion show that any theory starting in this region must be well defined in the UV.
\begin{center}
\begin{equation*}
\begin{tabular}{lll}
{{\hbox{\includegraphics[scale=0.85]{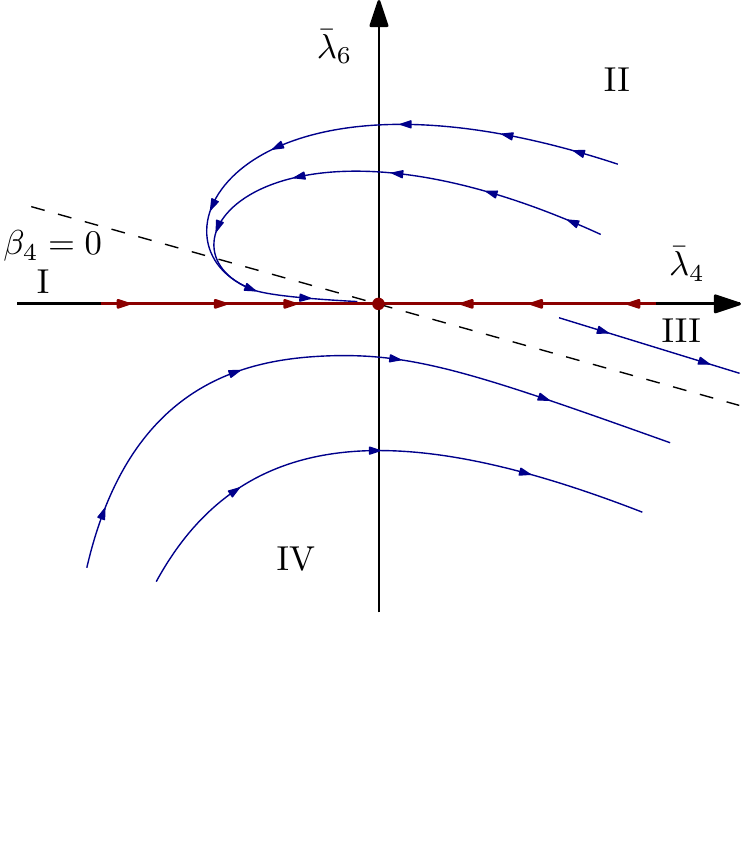} }}}&
{{\hbox{\includegraphics[scale=0.30]{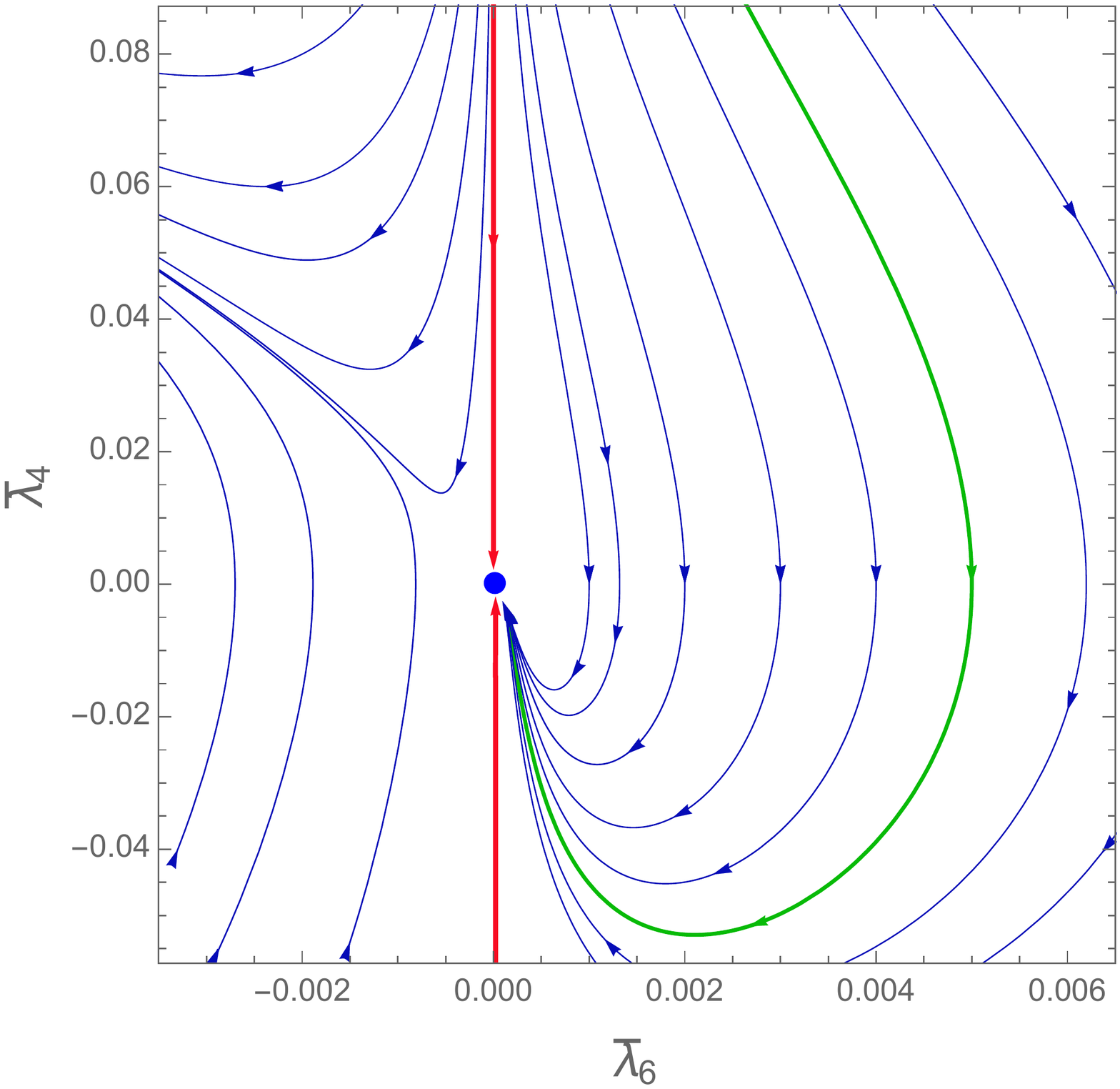} }}}\\
\qquad\qquad\qquad\qquad(a)&\qquad\qquad\qquad\qquad(b)&
\end{tabular}
\end{equation*}
\captionof{figure}{The qualitative picture of the RG trajectories in the plane $(\bar{\lambda}_4,\bar{\lambda}_6)$ (a). The numerical RG flow (b), where the green line corresponds to a typical trajectory on the upper plan, illustrating the discussion. }\label{RGGFP}
\end{center}
The additional term $18\Omega_3^2\bar{\lambda}_6$ in the equation \eqref{approxeta1} with respect to the one-loop approximation. Indeed, it introduce a strong attractive term in $\beta_6$, $-3\times18\Omega_3^2\bar{\lambda}_6^2$ coming from the contribution $-3\eta\bar{\lambda}_6$. Another attractive term $-2\times18\Omega_3^2\bar{\lambda}_6^2$ arise from the effective vertex $\overline{\pi_k^{(b_4^{(1)})}}$ ; so that the equation for $\beta_6$ is replaced by:
\begin{equation}
\beta_6\approx+\,6\Omega_3\bar{\lambda}_4\bar{\lambda}_6-5\times 18\Omega_3^2\bar{\lambda}_6^2 =6\Omega_3\bar{\lambda}_6 (\bar{\lambda}_4-15\Omega_3\bar{\lambda}_6) \,.
\end{equation}
This additional attractive term enhances the return of the flow lines toward the horizontal axis. The trajectories in the upper plan $\bar{\lambda}_6>0$ escape far away from this axis, and does not reach the Gaussian fixed point. Figure \ref{RGGFP1} represent the numerical RG trajectories in a small neighborhood of the Gaussian fixed point, taking into account this additional term.
\begin{center}
\begin{eqnarray*}
{\vcenter{\hbox{\includegraphics[scale=0.30]{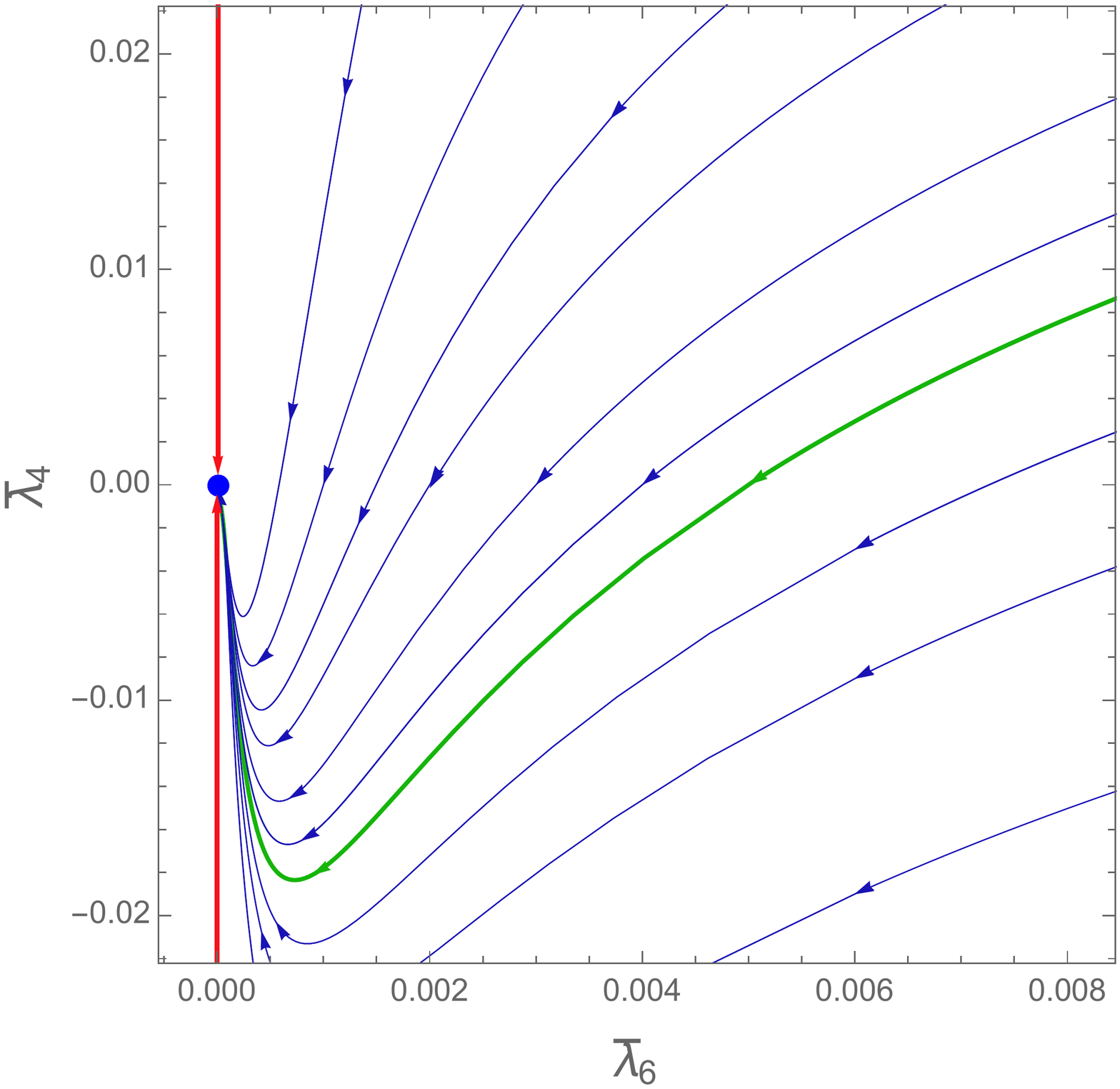} }}}
\end{eqnarray*}
\captionof{figure}{The qualitative picture of the RG trajectories in the plane $(\bar{\lambda}_4,\bar{\lambda}_6)$ On this figure, the same region of the phase space taking into account the additional term $18\Omega_3^2\bar{\lambda}_6$ in the equation \eqref{approxeta1}. The relative scales of the figures are chosen to optimise their visibility.}\label{RGGFP1}
\end{center}

\noindent
These conclusions seem to indicate that the expected UV-protected scenario in the positive region is lost. However, so far from the perturbative region, the behaviour of the RG trajectories remains unclear with the perturbative approach. A complementary analysis has to be performed using the full RG equation given by system \eqref{system21}-\eqref{system23}. Figure \ref{RGGFP2} provide a numerical integration of the RG equations in the plane $\bar{m}^2=0$ for different size of the neighbourhood surrounding the Gaussian fixed point. On Figure \ref{RGGFP2}a, in the vicinity of the Gaussian fixed point, we recover the landscape pictured on Figure \ref{RGGFP1}c. Furthermore, far away from the origin in the lower plan $\bar{\lambda}_6<0$, the trajectory seems to go away from the horizontal axis. Moreover, following the same arguments used to build Figure \ref{RGGFP}a, the incoming trajectories in the lower plan have no chance to reach the Gaussian fixed point and are repelled before to join him. Figure \ref{RGGFP2}b and \ref{RGGFP2}c seems to indicate the existence of an attractive vortex in the upper plan $\bar{\lambda}_6>0$. As for the one-loop analysis, the trajectory escaping from the Gaussian region return to him, attracted along a river reaching the Gaussian fixed point, after a long loop into the phase space. However, we have to be careful. For instance, we have completely neglected the role and the evolution of the mass parameter in this analysis. In the next section, we discuss the existence of a UV attractive fixed point, completing this picture of the RG flow.

\begin{center}
\begin{equation*}
\underset{a}{\vcenter{\hbox{\includegraphics[scale=0.4]{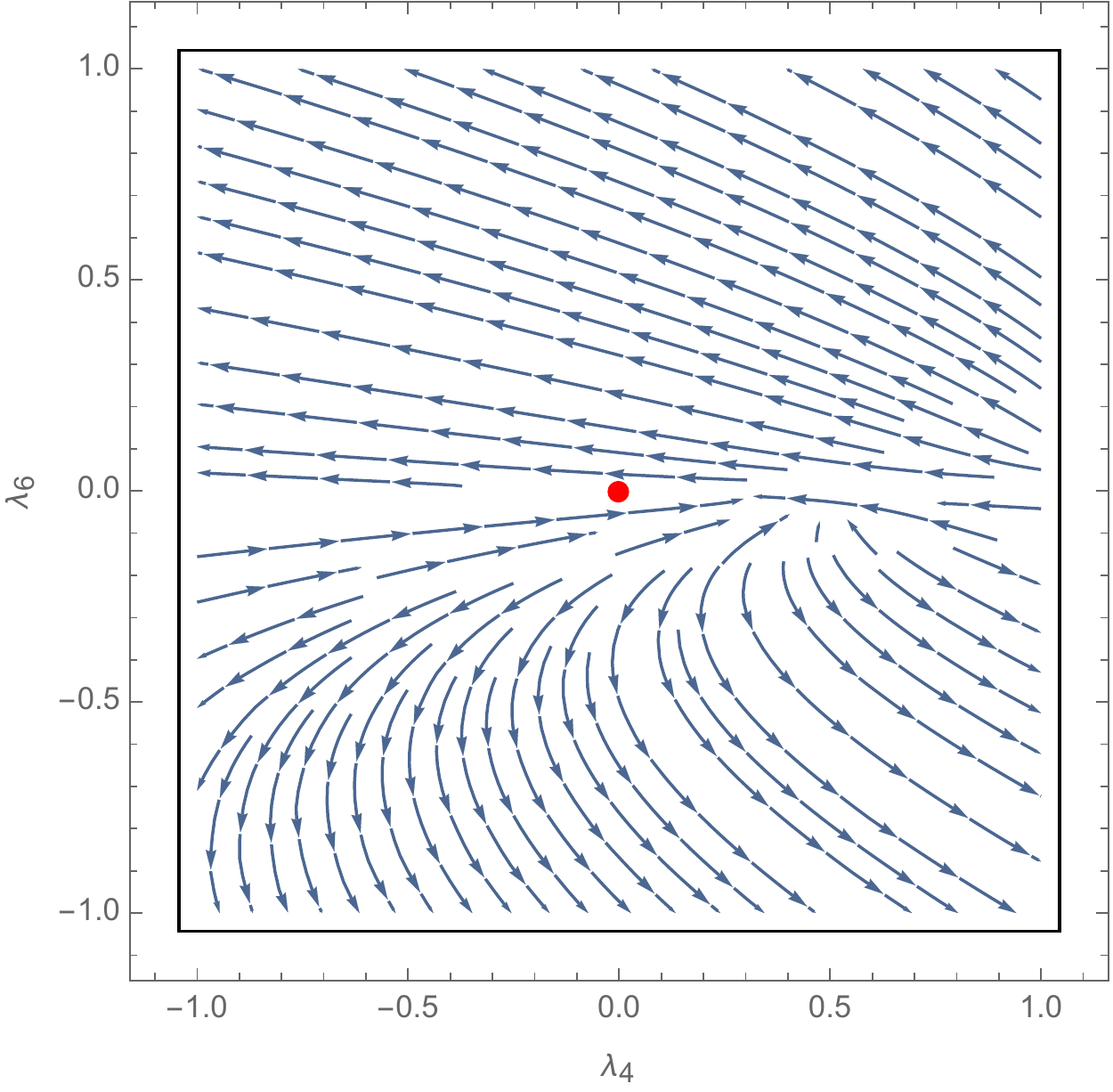} }}}\quad\underset{b}{\vcenter{\hbox{\includegraphics[scale=0.41]{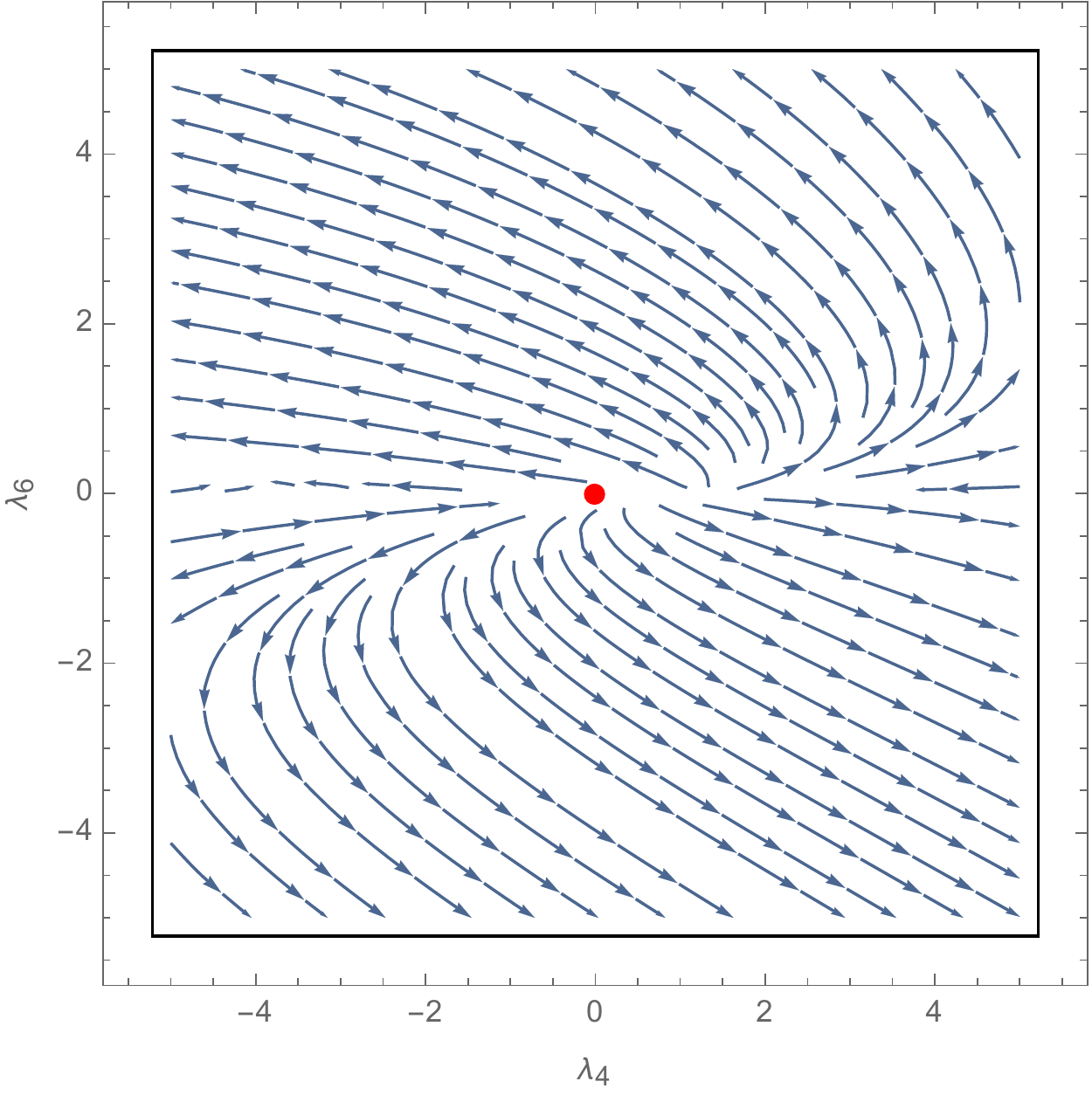} }}}\quad\underset{c}{\vcenter{\hbox{\includegraphics[scale=0.4]{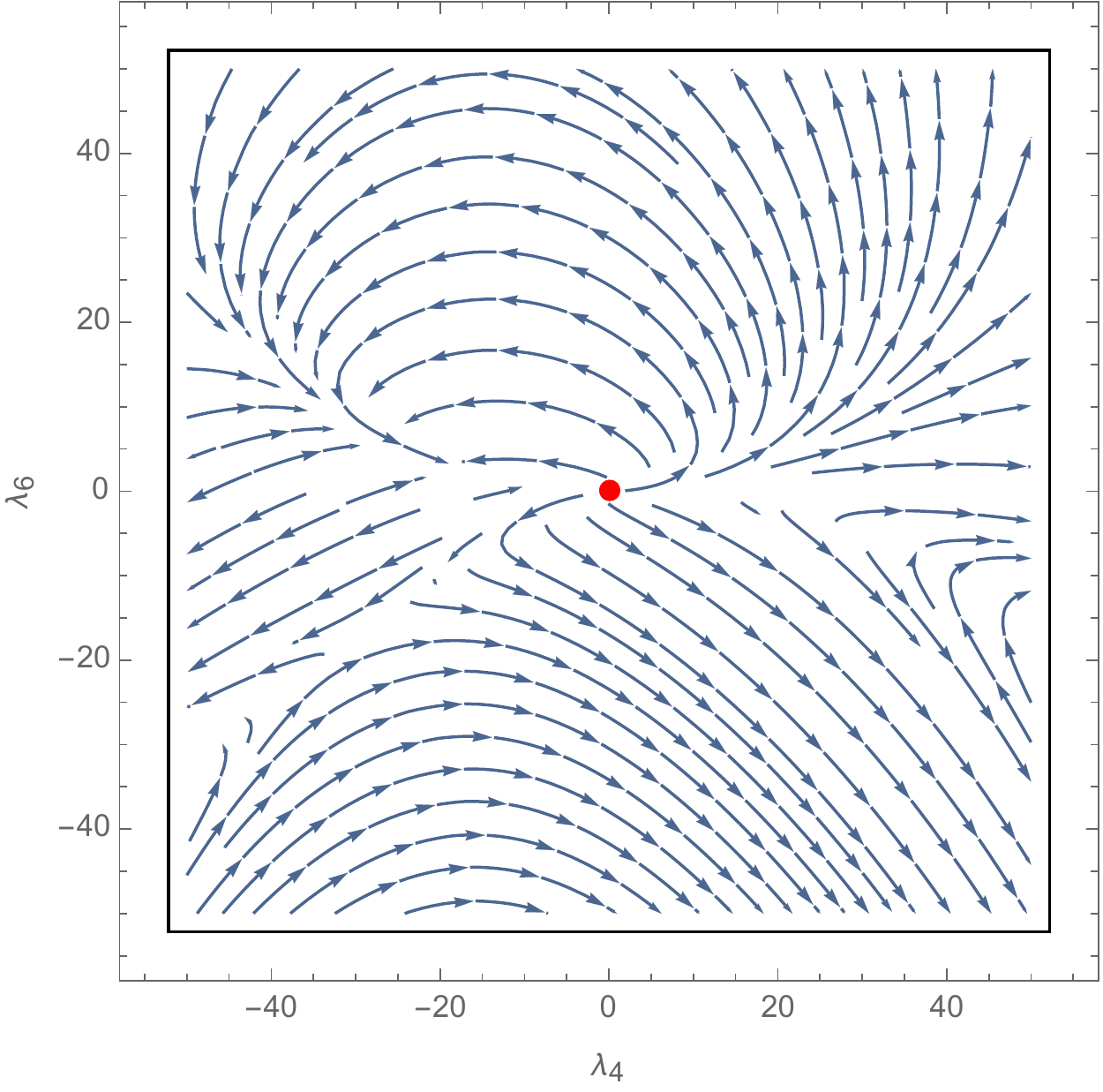} }}}
\end{equation*}
\captionof{figure}{The behavior of the numerical RG flow in the plan $\bar{m}^2=0$, from a very small region in the vicinity of the Gaussian fixed point (a) to a (relatively) large region (c). Again, the scale is adaptated to optimize the visibility.}\label{RGGFP2}
\end{center}
Investigating the solutions of the flow equations in the theory space we get the nontrivial fixed points \eqref{fp1},\eqref{fp2}, in which we can identified the Wilson-Fisher fixed point $FP_2$ which is previously announced in our introduction. None of these fixed points are compatible with the Ward identities.
\bea
\label{fp1}
FP_1&=&(\bar{m}^2\approx 0.019,\quad\bar{\lambda}_4\approx -0.0019,\quad\bar\lambda_6\approx -2.8\cdot 10^{-7})\\\label{fp2}
FP_2&=&(\bar m^2\approx -0.3701,\quad\bar\lambda_4\approx 0.34,\quad\bar\lambda_6\approx -0.367).
\eea
For $FP_1$ we get the following critical exponent: $(\theta_{11}=21.63,\theta_{12}=1.50,\theta_{13}=-0.81)$ which shows that this fixed point have two attractives directions and one repulsive direction in the UV. Note that the value of $\theta_{11}$ show how this attraction behave very fast in this UV limit. For $FP_2$ we get $(\theta_{2,1}=2.8+41.7 i,\theta_{22}=2.8-41.7i,\theta_{23}=-5.2+1.6i)$. Then the fixed point attracts the trajectories in two directions and the trajectory is repelled by the fixed point in one direction.
We will show in the next section that, in the physical subspace of the theory space all the nontrivial fixed points solution of the constraint flow equation disappeared in the domain with positive values of $\eta$.
\subsection{Ward-constrained phase-space in the deep UV}\label{sec5}
In addition to these flow equations \eqref{system21}, \eqref{system22},\eqref{system23}, Ward identities introduce a non-trivial constraint, which reduces the dimensionality of the effective flow, from $3$ to $2$. Indeed, on equation \eqref{approx3}, $Z_{-\infty}$ does not depends on the running scale $k$, implying that the right hand side has to be a constant along the flow. Differentiating the both sides with respect to $k$, this conservation may be translated locally as a constraint binding together the anomalous dimension $\eta$ and the beta functions $\{\beta_n\}$:
\begin{equation}
\eta\left(1-\frac{2\bar{\lambda}_4\Omega_3}{(1+\bar{m}^2)^{2}}\right)-\frac{2\Omega_3}{(1+\bar{m}^2)^{2}}\left(\beta_4-\frac{2\bar{\lambda}_4\beta_2}{1+\bar{m}^2}\right)+2\frac{1-\frac{2\bar{\lambda}_4\Omega_3}{(1+\bar{m}^2)^{2}}}{1-2\bar{\lambda}_4\bar{\mathcal{A}}_{k,2}}\left(\beta_4\bar{\mathcal{A}}_{k,2}-2\bar{\lambda}_4\bar{\mathcal{A}}_{k,3}\beta_2\right)=0\,,\label{constraint1}
\end{equation}
where we used of $\dot{\bar{\mathcal{A}}}_{k,2}=-2\bar{\mathcal{A}}_{k,3}$. For any fixed point $p=(\bar{m}^2_*,\bar{\lambda}_4^*,\bar{\lambda}_6^*)$, where the beta functions vanish, the previous relation split into two conditions:
\begin{equation}
\eta_*=0\,,\qquad \text{or}\, \qquad 1-\frac{2\bar{\lambda}_4^*\Omega_3}{(1+\bar{m}^2_*)^{2}}=0\,.
\end{equation}
To be physical, any fixed point of the system \eqref{system21}-\eqref{system23} has to be verify at least one of these two conditions. In order to get physical flow, we then have to project the RG equations given by the system \eqref{system21}-\eqref{system23} into the reduced phase space where the constraint \eqref{constraint1}. This is the aim of the next section \ref{sec4} to investigate this constrained physical phase space. However, before to move on this investigation, we close this section with some comments about our results. First of all, the constraint given by equation \eqref{sixpointWI} has been took into account in the derivation of $\eta$, equation \eqref{anomalousflow}. As a result, the flow equations given by the system \eqref{system21}-\eqref{system23} partially includes the constraint coming from the Ward identities. The constraint \eqref{approx3}, corresponding to the conservation of $Z_{-\infty}$ along the flow is the only missing information coming from the Ward identities in the $3$-dimensional parametric phase space building from the EVE method. The second comment is about the approximation used to compute the integrals. We used the LDE approximation, which is standard for TGFT in the symmetric phase for the computation of the flow equations. However, for the computation of the Ward constraint, and especially for the computation of the convergent integrals $\mathcal{A}_{k,j}$, we used of the same approximation, up to the hypothesis that the slowly decreasing of the denominator for large $\vec{p}\,^2$ may support the same approximation than the sharp cut-off coming from Heaviside functions. Consistency of the approximation has been checked in \cite{Lahoche:2019vzy}, but not for the same model. As a result, this additional approximation has to be into account in our final claims. \\

\noindent
Strong violations of the Ward identity \eqref{approx3} can be easily checked for several solutions of the flow equations \eqref{system21}-\eqref{system23} (see \cite{Lahoche:2018ggd} for an explicit violation concerning a rank $5$ TGFT). In this section we propose to take into account the constraint \eqref{constraint1} along with the flow following the strategy described in \cite{Lahoche:2019vzy}. As explained by the authors, the difficulty comes from the definition \eqref{defphi61} or \eqref{barphi6}, inherited from the structure of the melonic diagrams. This definition is too rigid to be conserved on the global phase space without introducing hard singularities. To describe the flow along the constrained subspace $\mathcal{E}$ along which \eqref{constraint1} is satisfied; we proceed as follows:\\

\noindent
(1) Keeping the equation \eqref{system21} for $\beta_2$, we replace the definition of $\beta_4$ provided by the flow equation by its expression coming from \eqref{constraint1}:

\begin{equation}
\beta_4=\frac{4 \beta_2 \bar\lambda_4\mathcal{X}(\bar m^2,\bar\lambda_4) +\eta (\bar{m}^2+1) (2 \bar{\mathcal{A}}_{k,2} \bar\lambda_4-1) \Big((\bar{m}^2+1)^2-2 \bar\lambda_4 \Omega_3\Big)}{2 \left(\bar{\mathcal{A}}_{k,2} (\bar{m}^2+1)^3-(\bar{m}^2+1) \Omega_3\right)}
\end{equation}
where
\bea
\mathcal{X}(\bar m^2,\bar\lambda_4):=\Omega_3 \Big(2 \bar{\mathcal{A}}_{k,2} \bar\lambda_4-2 \bar{\mathcal{A}}_{k,3} (\bar{m}^2+1) \bar\lambda_4-1\Big)+\bar{\mathcal{A}}_{k,3} (\bar{m}^2+1)^3
\eea
(2) Equaling this equation with the equation \eqref{system22}, we fix the value of $\bar\lambda_6$.

\noindent
(3) Substituting into equations \eqref{equationderiv} and \eqref{equationeta}, we get an explicit expression for the anomalous dimension over $\mathcal{E}$, replacing \eqref{anomalousflow}: $\eta_\mathcal{E}$. \\

\noindent
With this construction, the flow equations and the Ward identity are simultaneously verified along the RG trajectories. Moreover, as $\lambda_6$, the higher-order local observable like $\pi_k^{(b_4)}$ are fixed by the flow itself. $\pi_k^{(b_4)}$ is fixed by the equation \eqref{system23}, the left-hand side is explicitly computed from the explicit dynamical solution for $\lambda_6$, and so one. Our effective equation differs from the unconstrained ones \eqref{system21}-\eqref{system23}, except in the vicinity of the Gaussian fixed point.

\subsection{Numerical solution of the flow equations in the constraint theory space}
The numerical investigation of the flow equations providing from the EVE method (see \eqref{system21}, \eqref{system22} and \eqref{system23}) can be given carefully. First of all by considering the expression \eqref{system23} we can remark that the quantity $ \overline{\pi_k^{(b_4^{(1)})}}$ is specific from the EVE method. This expression does not exist by implementing the truncation as an approximation.
In this section, we discuss the solution of the new flow equations taking into account both the Ward constraint and the EVE, with the construction given in section \eqref{sec5}. Then we may compare this analysis with what follows in the last section. Finally, we discuss in detail the asymptotically freedom behaviour of this model.
After solving the Ward constraint flow equation numerically we get the Gaussian fixed point $p_0\approx (\bar{m}^2=0,\bar{\lambda}_4=0)$ and one new point $p_1\approx (\bar{m}^2=-0.36,\bar{\lambda}_4=0.018)$, and
the followings critical exponents and eigen-directions
\bea
&&p_0:\quad (\theta_{01},\theta_{02})\approx(2,1),\quad {\bf v}_{01}\approx(1,0),\quad {\bf v}_{02}\approx(-1,0.05),\quad \eta_{p_0}=0\\
&&p_1:\quad (\theta_{11},\theta_{12})\approx(1.3,-0.1),\quad {\bf v}_{11}\approx(1,0.034),\quad {\bf v}_{12}\approx(1,0.004),\quad\eta_{p_1}=0.10.
\eea
For $p_0$ the trajectory approaches the fixed points, and then is an UV attractive. In conclusion, our computation enforced our argument about the difficulty to find a physical fixed point compatible with the Ward identities appart the Gauxian fixed point which represents the trivial solution. Now let us discuss the behavior around the Gaussian fixed point and let try to show that the theory is asymptotically free. Using the constraint flow equation described above, we get asymptotically the following result around the Gaussian fixed point.
\bea
\left\{\begin{array}{cccc}
\beta_2&\approx-2\bar{m}^2-4\Omega_3 \bar{\lambda}_4\,,\\
\beta_4&\approx-\bar{\lambda}_4-6\Omega_3\bar{\lambda}_6\,,\\
\bar\lambda_6 &\approx-4\bar\lambda_4^2\\
\eta&\approx 8 \pi\bar\lambda_4,
\end{array}\right.
\eea
which is in agreement with the results obtained using only the EVE. Note that the $\beta$-function $\beta_4$ will be
\bea
\beta_4&\approx-\bar{\lambda}_4+24\Omega_3\bar\lambda_4^2+\mathcal{O}(\bar\lambda_4^2)
\eea
Then all values of the coupling tends to zero in the UV limit. The expression of $\beta_4=\beta_4^0\bar{\lambda}_4+\beta_4^1\bar{\lambda}_4^2$, and the sign of $\beta_4^0$ in the one loop perturbation computation, shows that the theory is asymptotically free in this UV limit i.e the flow is attractive around the Gaussian fixed point. The flow diagram around this fixed point is given in the figure \eqref{Figvincent23} from a numerical integration.
\begin{center}
\begin{equation*}
{\vcenter{\hbox{\includegraphics[scale=0.5]{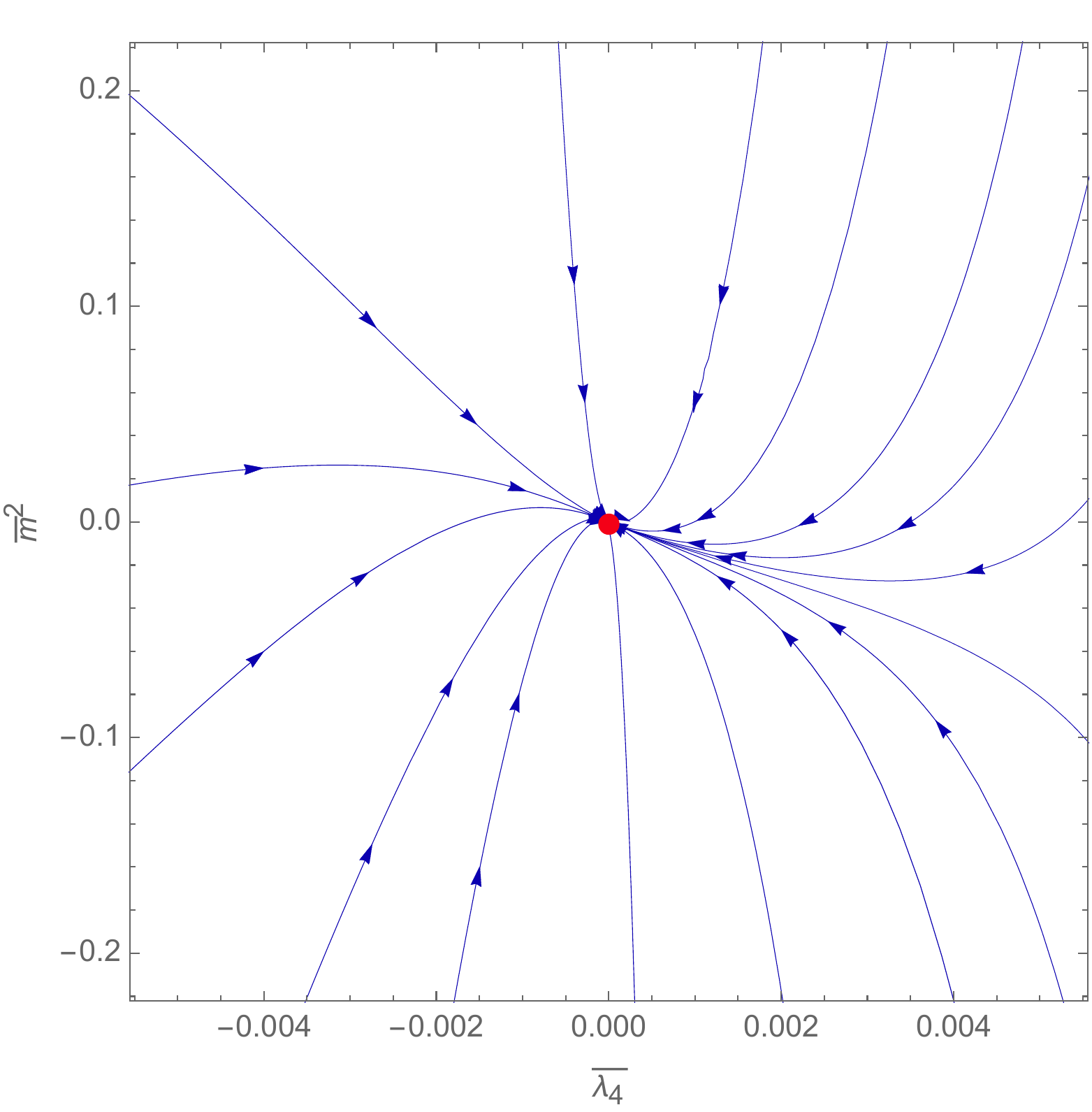} }}}
\end{equation*}
\captionof{figure}{The behavior of the numerical RG flow in the vicinity of the non-Gaussian fixed points in the space $\mathcal{E}$ }\label{Figvincent23}
\end{center}

Now consider the fixed point $p_1$ which have one attractive direction, one repulsive direction, and anomalous dimension $\eta^*=0.1$. This anomalous dimension enhances the canonical dimension, and following the definition of renormalized couplings, in the vicinity of the non-Gaussian fixed point, the $N$-point interactions scales with $k$ as $k^{d^*_N}$, with $d^*_N=d_N+\frac{N\eta^*}{2}$, $d_N=3-N/2$ being the canonical dimension, i.e. the scaling in the vicinity of the Gaussian fixed point. $d^*_N$ is negative for $N>6$, showing qualitatively that the list of relevant couplings is the same as in the perturbative regime. The values for the critical exponents are under this argument, in favour of the reliability of our conclusions.

\section{Conclusion}
In this work we build a solution to the nonperturbative functional renormalization group equation applied to a tensorial just renormalizable $\phi^6_4$ model, merging the effective vertex expansion and the Ward-Takahashi identities. In the strictly local potential approximation, these identities can be translated as a non-trivial relation between beta functions along with the flow. Focussing only on the non-branching sector, we prove asymptotic freedom (already conjectured in \cite{BenGeloun:2012yk}), as well as the existence of two non-Gaussian fixed points using the standard EVE method, and in particular the existence of a UV attractive fixed point matching with the heuristic arguments given in the introduction. These non-trivial solution of the renormalization group equation, however, are not compatible with the strong Ward constraint. Investigating the fully constrained space $\mathcal E$, connected to the Gaussian region, we discovered the existence of a new non-trivial fixed point, with one attractive and one repulsive direction. The characteristics of this fixed point seem to be following our approximations, in particular concerning the derivative expansion around marginal operators. Deep investigations have to be provided to improve this result, especially about interactions with disconnected boundaries, or derivative couplings, sorting out to the strictly local potential approximation.

\end{document}